\documentclass[aps,prb,twocolumn,superscriptaddress,showpacs,amsmath,amssymb,footinbib,longbibliography,10pt]{revtex4-2}
\usepackage[utf8]{inputenc}
\usepackage{bm}
\usepackage[usenames]{color}
\usepackage{multirow}
\usepackage{amssymb}
\usepackage{amsbsy}
\usepackage{amsmath}
\usepackage{stmaryrd}
\usepackage{graphicx}
\usepackage{epsfig}
\usepackage{placeins}
\usepackage[normalem]{ulem}
\usepackage{bbold}
\usepackage{bbm}
\usepackage{braket}
\usepackage{graphicx}
\usepackage{tikz}
\usepackage{color}
\usetikzlibrary{calc}
\usetikzlibrary{patterns}
\usepackage[colorlinks,linkcolor=blue,citecolor=blue,urlcolor=blue]{hyperref}

\makeatletter

\setcitestyle{numbers,square,comma,sort&compress}

\nocite{*}

\definecolor{palette1}{HTML}{A8216B}
\definecolor{palette2}{HTML}{F1184C}
\definecolor{palette3}{HTML}{F36943}
\definecolor{palette4}{HTML}{F7DC66}
\definecolor{palette5}{HTML}{2E9599}

\usepackage{scalerel}
\usepackage{tikz}
\usetikzlibrary{calc}
\usetikzlibrary{patterns}
\usetikzlibrary{svg.path}
\definecolor{orcidlogocol}{HTML}{A6CE39}
\tikzset{
  orcidlogo/.pic={
    \fill[orcidlogocol]
svg{M256,128c0,70.7-57.3,128-128,128C57.3,256,0,198.7,0,128C0,57.3,57.3,0,128,
0C198.7,0,256,57.3,256,128z};
    \fill[white] svg{M86.3,186.2H70.9V79.1h15.4v48.4V186.2z}

svg{M108.9,79.1h41.6c39.6,0,57,28.3,57,53.6c0,27.5-21.5,53.6-56.8,53.6h-41.8V79.
1z
M124.3,172.4h24.5c34.9,0,42.9-26.5,42.9-39.7c0-21.5-13.7-39.7-43.7-39.7h-23.
7V172.4z}

svg{M88.7,56.8c0,5.5-4.5,10.1-10.1,10.1c-5.6,0-10.1-4.6-10.1-10.1c0-5.6,4.5-10.1
,10.1-10.1C84.2,46.7,88.7,51.3,88.7,56.8z};
  }
}

\newcommand\orcid[1]{\!%
  \href{https://orcid.org/#1}{%
    \mbox{%
      \scaleto{%
        \begin{tikzpicture}[yscale=-1,transform shape]
          \pic{orcidlogo};
        \end{tikzpicture}
      }{8pt}%
    }%
  }%
}

\begin{document}
\title{Lindblad dynamics from spatio-temporal correlation functions in
nonintegrable spin-$1/2$ chains with different boundary conditions}

\author{Markus Kraft~\orcid{0009-0008-4711-5549}}
\email{markus.kraft@uos.de}
\affiliation{University of Osnabr{\"u}ck, Department of Mathematics/Computer
Science/Physics, D-49076 Osnabr{\"u}ck, Germany}

\author{Jonas Richter~\orcid{0000-0003-2184-5275}}
\affiliation{Department of Physics, Stanford University, Stanford, CA 94305,
USA}
\affiliation{Institut f{\"u}r Theoretische Physik, Leibniz Universit\"at
Hannover, 30167 Hannover, Germany}

\author{Fengping Jin~\orcid{0000-0003-3476-524X}}
\affiliation{Institute for Advanced Simulation, J\"ulich Supercomputing Centre,
Forschungszentrum J\"ulich, D-52425 J\"ulich, Germany}

\author{Sourav Nandy}
\affiliation{Jo\v zef Stefan Institute, SI-1000 Ljubljana, Slovenia}

\author{Jacek Herbrych~\orcid{0000-0001-9860-2146}}
\affiliation{Wroclaw University of Science and Technology, 50-370 Wroclaw,
Poland}

\author{Kristel Michielsen~\orcid{0000-0003-1444-4262}}
\affiliation{Institute for Advanced Simulation, J\"ulich Supercomputing Centre,
Forschungszentrum J\"ulich, D-52425 J\"ulich, Germany}

\author{Hans De Raedt~\orcid{0000-0001-8461-4015}}
\affiliation{Zernike Institute for Advanced Materials, University of Groningen,
NL-9747 AG Groningen, Netherlands}

\author{Jochen Gemmer}
\affiliation{University of Osnabr{\"u}ck, Department of Mathematics/Computer
Science/Physics, D-49076 Osnabr{\"u}ck, Germany}

\author{Robin Steinigeweg~\orcid{0000-0003-0608-0884}}
\email{rsteinig@uos.de}
\affiliation{University of Osnabr{\"u}ck, Department of Mathematics/Computer
Science/Physics, D-49076 Osnabr{\"u}ck, Germany}

\date{\today}


\begin{abstract}
We investigate the Lindblad equation in the context of boundary-driven
magnetization transport in spin-$1/2$ chains. Our central question is whether
the nonequilibrium steady state of the open system, including its buildup in
time, can be described on the basis of the dynamics in the closed system. To
this end, we rely on a previous work [Phys.\ Rev.\ B {\bf 108}, L201119
(2023)], where a description in terms of spatio-temporal correlation functions
has been suggested in the case of weak driving and small system-bath coupling.
Because this work has focused on integrable systems and periodic boundary
conditions, we here extend the analysis in three directions: We (i) consider
nonintegrable systems, (ii) take into account open boundary conditions and
other bath-coupling geometries, and (iii) provide a comparison to time-evolving
block decimation. While we find that nonintegrability plays a minor role, the
choice of the specific boundary conditions can be crucial, due to potentially
nondecaying edge modes. Our large-scale numerical simulations suggest that a
description based on closed-system correlation functions is an useful
alternative to already existing state-of-the-art approaches.
\end{abstract}

\maketitle


\section{Introduction}

Quantum many-body systems out of equilibrium are a central topic of modern
physics, and they have attracted increasing attention over recent years, both
experimentally and theoretically \cite{Bloch2008, Abanin2019, Polkovnikov2011,
Eisert2015, Dalessio2016}. Key questions in this context are the emergence and
properties of steady states in the limit of long times, but also the actual
route to such states in the course of time \cite{Abanin2019, Polkovnikov2011,
Eisert2015, Dalessio2016}. The understanding of these questions is of
importance to isolated systems without any coupling to an environment and open
systems with a weak or strong coupling to a bath, and it has witnessed a rather
remarkable progress due to fresh concepts like eigenstate thermalization
\cite{Deutsch1991, Srednicki1994, Rigol2008} and the typicality of random pure
states \cite{Bartsch2009, Gemmer2004, Goldstein2006, Popescu2006, Reimann2007,
Elsayed2013, Steinigeweg2014, Heitmann2020, Jin2021} as well as due to the
development of sophisticated numerical techniques \cite{Schollwoeck2005,
Schollwoeck2011}.

In systems with a globally conserved quantity, a quite natural nonequilibrium
process is given by transport \cite{Bertini2021}. It is particularly a prime
example of relevance to closed and open systems alike, in addition to the
relevance of steady states and the relaxation to them. In isolated systems, a
widely used approach is linear-response theory, which yields the well-known
Kubo formula in terms of current autocorrelation functions \cite{Kubo2012}.
This theory can be formulated for density-density correlation functions as
well, either in momentum and frequency space, or in the space and time domain.
While linear response provides a clear-cut strategy, the concrete evaluation of
correlation functions for specific models has turned out to be an analytical
and numerical challenge, even for seemingly simple models of Heisenberg or
Hubbard type in one dimension \cite{Bertini2021}, with the most recent progress
by generalized hydrodynamics \cite{Bastianello2022, Doyon2023}.

In an open-system scenario, transport can be induced by the coupling of the
system to baths at different temperatures or chemical potentials, which then
usually yields a nonequilibrium steady state (NESS) with a constant current and
characteristic density profile in the long-time limit \cite{Prosen2009,
Michel2003, Wichterich2007, Znidaric2011}. Such a scenario is often modelled by
an equation of Lindblad form \cite{Breuer2007}. While the derivation of this
equation from a microscopic system-bath model is a nontrivial task in practice
\cite{Wichterich2007, DeRaedt2017}, it is the most general version of a
time-local master equation, which maps any density matrix to a density matrix
again. In particular, it allows for an efficient numerical treatment by
matrix-product states for quite large system sizes \cite{Prosen2009,
Verstraete2004, Zwolak2004, Weimer2021}, as dissipation reduces the unavoidable
growth of entanglement as a function of time.

The two complementary approaches of closed and open systems have been the
basis for a reliable identification of different dynamical phases, including
the case of normal diffusion \cite{Bertini2021}, but also for other transport
\cite{Prosen2011, Prosen2013, Ljubotina2019, Gopalakrishnan2019a,
Bulchandani2021, Nandy2023, Serbyn2021, Singh2021, Richter2022a, Richter2022,
Nandkishore2015, Luitz2017a} types ranging from ballistic to anomalous dynamics.
Moreover, quantitative agreement of diffusion constants has been found in some
cases \cite{Steinigeweg2009a, Steinigeweg2009c, Znidaric2018, Znidaric2019}.
However, a clear correspondence between the dynamics in closed and open systems
is still lacking \cite{Kundu2009, Purkayastha2018, Purkayastha2019}. Thus, the
main question of the present paper is: Can one predict the open-system dynamics
based on the knowledge of the closed-system time evolution? While a general
answer to this question appears to be hopeless, it might be possible for
specific models and parameters at least.

\begin{figure}[t]
\includegraphics[width=0.9\columnwidth]{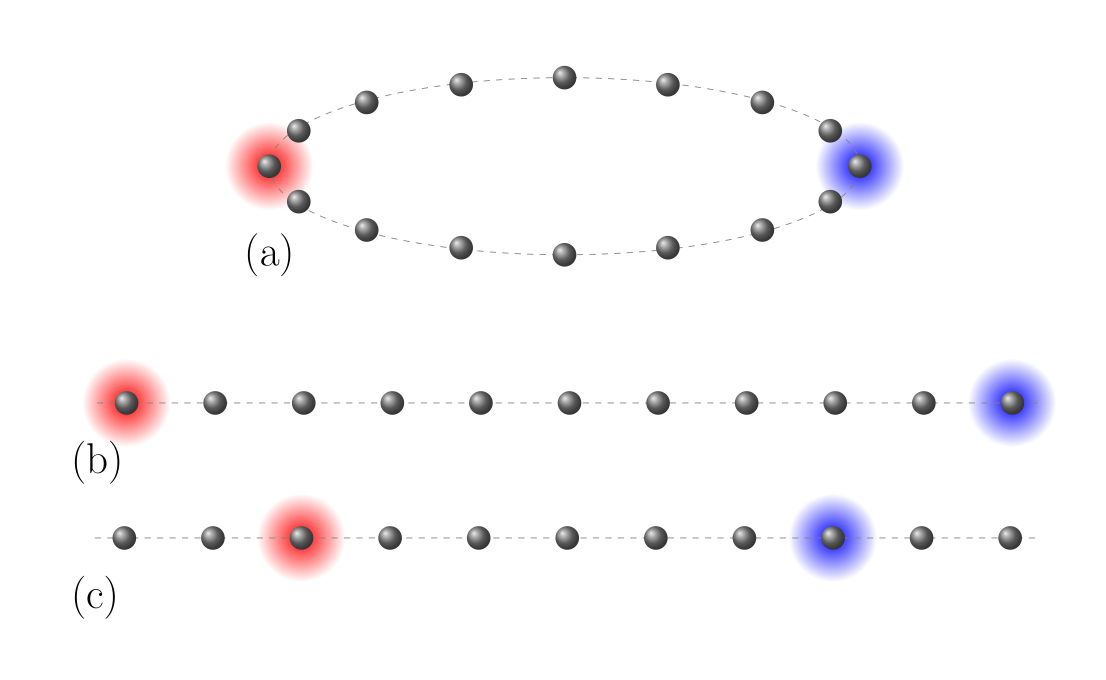}
\caption{Sketch of the three different geometries of the bath coupling. (a)
Periodic boundary conditions. (b) and (c) Open boundary conditions with a bath
coupling located exactly at the edges or close to the edges of the system. Note
that open boundary conditions should not be confused with open-system scenario.}
\label{fig:sketch1}
\end{figure}

To investigate this question, we rely here on a previous work
\cite{Heitmann2023}, where spatio-temporal correlation functions have been
suggested as a convenient ingredient in the case of weak driving and small
system-bath coupling. Because the aforementioned work has focused on integrable
systems and periodic boundary conditions, we intend to extend the analysis in
three different directions: We (i) consider nonintegrable systems, (ii) take
into account open boundary conditions and other bath-coupling geometries, see
Fig.\ \ref{fig:sketch1}, and (iii) provide a comparison to time-evolving block
decimation (TEBD). While we find that nonintegrability plays a minor role, the
choice of the specific boundary conditions can be crucial, due to potentially
nondecaying edge modes. Our large-scale numerical simulations suggest that a
description based on closed-system correlation functions constitutes an useful
alternative to existing state-of-the-art approaches.

Our paper is organized as follows. To begin with, we introduce in Sec.\
\ref{sec:closed} the closed-system models studied and the spatio-temporal
correlation functions. Then, in Sec.\ \ref{sec:DQT}, we discuss the concept of
dynamical quantum typicality, and describe its implications for numerical and
analytical purposes. Afterwards, in Sec.\ \ref{sec:open}, we continue with the
open-system setup, before we review the technique of stochastic unraveling in
Sec.\ \ref{sec:SU}. The subsequent Sec.\ \ref{sec:connection} is devoted to the
central prediction used later, and its underlying assumptions. Eventually, in
the next Sec.\ \ref{sec:results}, we present our results. We conclude in Sec.\
\ref{sec:conslusion}, and give additional information in the appendix.


\section{Closed models and spatio-temporal correlation functions}
\label{sec:closed}

In this paper, we consider two different paradigmatic examples for quantum
many-body models, which have attracted significant attention in the literature
on, e.g., transport before. These two examples are nonintegrable modifications
of the integrable spin-$1/2$ XXZ model in one dimension. The Hamiltonian of
this model is given by \cite{Bertini2021}
\begin{equation} \label{eq:XXZ}
H_\text{obc} = J \sum_{r=1}^{N-1} (S_{r}^{x} S_{r+1}^{x} + S_{r}^{y}
S_{r+1}^{y} + \Delta S_{r}^{z} S_{r+1}^{z}) \, ,
\end{equation}
where $S_{r}^{i}$ $(i = x,y,z)$ are spin-$1/2$ operators at site $r$, $N$
is the total number sites, $J > 0$ is the antiferromagnetic exchange coupling
constant, and $\Delta$ is the anisotropy in $z$ direction. While the
Hamiltonian in Eq.\ (\ref{eq:XXZ}) is denoted for open boundary conditions, we
will also use periodic boundary conditions,
\begin{equation}
H_\text{pbc} = H_\text{obc} + J (S_{N}^{x} S_{1}^{x} + S_{N}^{y} S_{1}^{y} +
\Delta S_{N}^{z} S_{1}^{z}) \, ,
\end{equation}
where the numbers of sites $N$ is chosen to be even.
The specific choice of boundary conditions will play an important role for
the open setup, which will be discussed later in detail.

The spin-1/2 XXZ chain is well-known to be integrable for any value of
$\Delta$, and it has been in the focus of our previous work
\cite{Heitmann2023}. In this work, we go beyond and include
integrability-breaking perturbations, which then yield a more generic
situation. As a first type of perturbation, we choose further interactions
between next-to-nearest sites, which lead for open boundary conditions to
\begin{equation} \label{eq:perturbation1}
H'_\text{obc} = H_\text{obc} + J \Delta' \sum_{r=1}^{N-2} S_{r}^{z} S_{r+2}^{z}
\end{equation}
and, although not shown explicitly, to a corresponding form for periodic
boundary conditions. Here, $\Delta'$ is the strength of the perturbation, and
we will focus on the particular value $\Delta' = 0.5$, for which integrability
is well broken. As a second type of perturbation, we choose a Zeeman term with
a staggered magnetic field,
\begin{equation} \label{eq:perturbation2}
H''_\text{obc/pbc} = H_\text{obc/pbc} + B \sum_{r=1}^N (-1)^r S_r^z \, ,
\end{equation}
where we will set the strength of the perturbation to the specific value $B/J =
0.5$, for the same reason as the one above.

\begin{figure}[t]
\includegraphics[width=0.9\columnwidth]{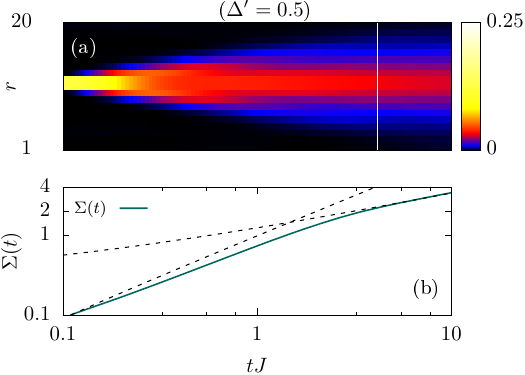}
\caption{(a) Time-space density plot of the correlation function $\langle
S_{r}^{z}(t) S_{r'}^{z}(0)\rangle_{\text{eq}}$, as obtained numerically for the
Hamiltonian $H'$ in Eq.\ (\ref{eq:perturbation1}) with $\Delta=1.5$,
$\Delta'=0.5$, $N=20$, $r'=N/2+1=11$, and periodic boundary conditions. (b) Corresponding time
evolution of the standard deviation $\Sigma(t)$ as well as two power laws
$\propto t^1$ and $\propto t^{1/2}$, in a double-logarithmic plot.}
\label{fig:sketch2}
\end{figure}

Since the total magnetization $S^z = \sum_r S_r^z$ is strictly conserved for the
models in Eqs.\ (\ref{eq:perturbation1}) and (\ref{eq:perturbation2}), $[H,
S^z] = 0$, transport of spins is a meaningful question. Within the different
approaches to transport in general, linear response theory is one of the main
concepts. While this theory leads to the Kubo formula and current
autocorrelation functions \cite{Kubo2012}, it is also the basis for
spatio-temporal correlation functions,
\begin{equation} \label{eq:correlations}
\langle S_{r}^{z}(t) S_{r'}^{z}(0) \rangle_{\text{eq}} =
\frac{\text{tr} [ e^{-\beta H} e^{\mathrm i H t} S_{r}^{z} e^{-\mathrm i H t}
S_{r'}^{z} ] }{\text{tr}[e^{-\beta H}]} \ .
\end{equation}
Here, $\beta = 1/T$ is the inverse temperature (measured in the units of the
Boltzmann constant), and from now on we will consider the high-temperature
limit $\beta \to 0$, which still features nontrivial transport properties. The
correlation functions in Eq.\ (\ref{eq:correlations}) measure the overlap of a
time-evolved $S_r^z(t)$ at some site $r$ with an initial $S_{r'}^z(0)$ at
another site $r'$. In Fig.\ \ref{fig:sketch2} (a), we illustrate the space-time
dependence, as obtained numerically for the Hamiltonian $H'$ in Eq.\
(\ref{eq:perturbation1}) with $\Delta=1.5$, $\Delta'=0.5$, $N=20$, and periodic
boundary conditions. Initially, there is a $\delta$-function peak at the site
$r = r'$ and an uniform equilibrium background at other sites $r \neq r'$. The
subsequent broadening of the $\delta$-function peak corresponds to transport.

A convenient way to analyze the type of transport is provided by the
spatial variance \cite{Steinigeweg2009b}
\begin{equation} \label{eq:variance}
\Sigma^{2}(t)= \sum_{r}(r-r')^{2} C_{rr'}(t) - \left[\sum_{r}(r-r') C_{rr'}(t)
\right]^{2} \, ,
\end{equation}
where the distribution $C_{rr'}(t) = 4 \langle
S_{r}^{z}(t)S_{r'}^{z}\rangle_{\text{eq}}$ is a normalized version of Eq.\
(\ref{eq:correlations}), $\sum_r C_{rr'}(t) = 1$. For diffusive transport,
$\Sigma(t) \propto t^{1/2}$. In Fig.\ \ref{fig:sketch2} (b), we depict
$\Sigma(t)$ for the example in Fig.\ \ref{fig:sketch2} (a). While it is clear
that a ballistic growth $\Sigma(t) \propto t$ has to take place at short times,
there is a diffusive growth $\Sigma(t) \propto t^{1/2}$ at intermediate times,
as expected due to the nonintegrability of the model \cite{note1}. It is worth
mentioning that the integrable model in Eq.\ (\ref{eq:XXZ}) features a richer
phase diagram, including superdiffusive behavior for $\Delta = 1$ and ballistic
behavior for $\Delta < 1$. The different types of transport become also manifest
in the quantity \cite{Steinigeweg2009b}
\begin{eqnarray}
D(t) = \frac{1}{2} \frac{\text{d}}{\text{d}t} \Sigma^{2}(t) \, ,
\end{eqnarray}
which becomes constant in case of diffusion.

Let us already mention here that the spatio-temporal correlation function
$\langle S_{r}^{z}(t) S_{r'}^{z}(0) \rangle_{\text{eq}}$ is not only
a strategy to study transport in closed systems, but it may also be used to
predict the build-up of a nonequilibrium steady state in open systems,
where a bath is coupled at each edge. Investigating the quality of such
a prediction is a central point of our paper. However, before we discuss this
point in detail, we need to introduce further concepts in the following.

\section{Dynamical quantum typicality} \label{sec:DQT}

Next, let us discuss dynamical quantum typicality as one of the central
concepts applied in this paper. On the one hand, this concept provides the basis
for a numerical calculation of the spatio-temporal correlation functions
$\langle S_{r}^{z}(t) S_{r'}^{z}(0)\rangle_{\text{eq}}$ in closed systems of
comparatively large size. On the other hand, it constitutes a main ingredient
to connect these correlation functions to the dynamics in open systems, as we
will see in a later section.

Loosely speaking, the basic idea of typicality is that a single pure state
can imitate the full statistical ensemble, on the level of the corresponding
expectation values \cite{Gemmer2004, Goldstein2006, Popescu2006, Reimann2007,
Bartsch2009}. To be precise, we introduce a pure state drawn at random
(according to the Haar measure) from a Hilbert space of high dimension $D$,
\begin{equation} \label{eq:random_state}
| \psi \rangle = \sum_{n=1}^{D} c_{j} \, | j \rangle \, ,
\end{equation}
where $\{ | j \rangle \}$ is an arbitrary orthonormal basis, and the real and
imaginary part of the coefficients $c_{j} = a_{j} + i b_{j}$ result from
a Gaussian probability distribution with zero mean and unit variance. For such a
pure state, one then obtains the approximation \cite{Heitmann2020, Jin2021}
\begin{equation}
\langle S_{r}^{z}(t) S_{r'}^{z}(0) \rangle_{\text{eq}} = \frac{\langle \psi
| S_{r}^{z}(t) S_{r'}^{z}(0) | \psi \rangle}{\langle \psi | \psi \rangle} +
{\cal O} \Big (\frac{1}{\sqrt{D}} \Big ) \, ,
\end{equation}
where the statistical error on the r.h.s.\ is exponentially small in
system size, due to $D = 2^N$. This approximation can be rewritten as
\begin{equation} \label{eq:two_states}
\langle S_{r}^{z}(t) S_{r'}^{z}(0) \rangle_{\text{eq}} \approx \frac{\langle
\psi(t) | S_{r}^{z} | \phi(t) \rangle}{\langle \psi | \psi \rangle} \, ,
\end{equation}
using the two auxiliary pure states $|\psi(t) \rangle = e^{-i H
t}| \psi \rangle$ and $|\phi(t) \rangle = e^{-i H t} S_{r'}^z | \psi \rangle$.
The expression (\ref{eq:two_states}), just like analogous expressions for other
observables, has turned out to be particularly useful for numerical simulations,
since its evaluation requires the forward propagation of pure states in time.
These propagations can be carried out efficiently in huge Hilbert spaces, which
are orders of magnitude larger than the ones accessible by standard exact
diagonalization \cite{Heitmann2020, Jin2021}. Note that the numerical data in
Fig.\ \ref{fig:sketch2} are also obtained in this way, yet on the basis of a
slightly different and simpler expression, as explained in the following.

The simplification employs the fact that $S_r^z$, as well as $n_r = S_r^z +
1/2$, are operators with the specific properties $\text{tr} [S_r^z] = 0$ and
$n_r^2 = n_r$. Using these two properties, and introducing the pure state
\begin{equation} \label{eq:L1_state}
|\varphi(t) \rangle = e^{-i H t} n_{r'} | \psi \rangle \, ,
\end{equation}
then leads to expression \cite{Steinigeweg2017a}
\begin{equation} \label{eq:one_state}
\langle S_{r}^{z}(t) S_{r'}^{z}(0) \rangle_{\text{eq}} \approx
\frac{1}{2} \frac{\langle
\varphi(t) | S_{r}^{z} | \varphi(t) \rangle}{\langle \varphi | \varphi \rangle}
\, ,
\end{equation}
which involves a single pure state only. This expression is an obvious
numerical benefit, but also provides a central analytical relation for later
purposes in the context of open systems.


\section{Open setup and Lindblad equation} \label{sec:open}

Now, let us turn to an open-system scenario, where we couple the system to an
environment. We describe this scenario by the Lindblad equation,
\begin{equation} \label{eq:Lindblad}
\dot{\rho}(t) = \mathcal{L}\rho(t) = \mathrm i [\rho(t),H] + \mathcal{D}
\rho(t) \, ,
\end{equation}
as the most general form of a time-local quantum master equation,
which maps any density matrix to a density matrix again
\cite{Breuer2007}. The Lindblad equation
(\ref{eq:Lindblad}) consists of a coherent part for the unitary time evolution
w.r.t.\ $H$ and an incoherent damping term. This damping term is given by
\begin{equation}
\mathcal{D} \rho(t) = \sum_{j} \alpha_{j} \left( L_{j}\rho(t)L_{j}^{\dagger} -
\frac{1}{2} \left \{ \rho(t) , L_{j}^{\dagger} L_{j} \right \} \right) \, ,
\end{equation}
with non-negative rates $\alpha_{j}$, Lindblad operators $L_{j}$, and the
anticommutator $\{ \bullet,\bullet \}$. Despite the generality of the Lindblad
equation, its derivation is a challenging task for a specific microscopic model
\cite{Wichterich2007, DeRaedt2017}.

Here, we couple our system to two baths and choose the respective Lindblad
operators \cite{Bertini2021}
\begin{eqnarray}
L_{1} = S^{+}_{B_{1}} \, , & \alpha_{1} = \gamma(1+\mu) \, , \label{eq:L1} \\
L_{2} = L^{\dagger}_{1} = S^{-}_{B_{1}} \, , & \alpha_{2} = \gamma(1-\mu) \, ,
\\
L_{3} = S^{+}_{B_{2}} \, , & \alpha_{3} = \gamma(1-\mu) \, , \\
L_{4} = L^{\dagger}_{3} = S^{-}_{B_{2}} \, , & \alpha_{4} = \gamma(1+\mu) \, ,
\label{eq:L4}
\end{eqnarray}
where $\gamma$ is the system-bath coupling and $\mu$ is the driving strength.
$L_{1}$ and $L_{2}$ are local operators, which act on a site $B_{1}$ and flip a
spin up and down, respectively. $L_{3}$ and $L_{4}$ are corresponding operators
at another site $B_{2}$.

Our different choices of the bath-contact sites $B_1$ and $B_2$ are illustrated
in Fig.\ \ref{fig:sketch1}. In case of periodic boundary conditions, we set
$B_1 = 1$ and $B_{2} = N/2 + 1$ in a distance $N/2$, see Fig.\
\ref{fig:sketch1} (a). In case of open boundary conditions, we set $B_1 = 1$ at
the left edge and $B_2 = N$ at the right edge or, as an alternative, $B_1$ and
$B_2$ close to the edges, see Figs.\ \ref{fig:sketch1} (b) and (c). This
alternative will be discussed in detail later. For all choices of $B_1$ and
$B_2$, the first (second) bath induces a net polarisation of order $\mu$
($-\mu$), leading to a steady state in the long-time limit with a characteristic
density profile and a constant current.

In this open scenario, we are interested in the dynamics of local
magnetization, as occurring at finite times and in the long-time limit. Thus, we
study expectation values
\begin{equation}
\langle S_{r}^{z}(t) \rangle = \text{tr}[ \rho(t) S_{r}^{z} ]\ ,
\label{local_magnetization_expatation_value_Lindblad}
\end{equation}
which depend on the parameters of the Hamiltonian $H$, but also on the two bath
parameters $\mu$ and $\gamma$. As initial condition, we choose $\rho(0) \propto
1$, which corresponds to the high-temperature limit $\beta \to 0$ and a
homogeneous profile of magnetization.


\section{Stochastic unraveling of the Lindblad equation} \label{sec:SU}

We aim at finding a solution of the Lindblad equation or an accurate
approximation of the same. To this end, we rely on the concept of stochastic
unraveling, which uses pure states $|\psi \rangle$ rather than density matrices
$\rho$ \cite{Dalibard1992, Michel2008}. This concept consists of an alternating
sequence of stochastic jumps with one of the Lindblad operators and, between
the stochastic jumps, a deterministic time evolution w.r.t.\ an effective
Hamiltonian,
\begin{equation}
H_{\text{eff}} = H - \frac{i}{2} \sum_{j} \alpha_{j} L_{j}^{\dagger} L_{j}
\, .
\end{equation}
For our choice of Lindblad operators in Eqs.\ (\ref{eq:L1}) - (\ref{eq:L4}),
this effective Hamiltonian takes on the form
\begin{equation}
H_{\text{eff}} = H - i \gamma + i \gamma \mu (n_{B_{1}} - n_{B_{2}})
\end{equation}
with the occupation number $n_{r} = S_{r}^{+}S_{r}^{-} = S_{r}^{z} +
1/2$. In this work, we focus on the weak-driving case. For $\mu \ll 1$,
$H_{\text{eff}}$ can be approximated as
\begin{equation}
H_{\text{eff}} \approx H - i \gamma \, .
\end{equation}
Hence, the time evolution w.r.t.\ to $H_{\text{eff}}$ becomes
\begin{eqnarray} \label{eq:decay}
|\psi (t) \rangle \approx e^{-\gamma t} \, e^{-i H t} \, |\psi \rangle \, .
\end{eqnarray}
Therefore, the dynamics of a pure state is only generated by the closed system
$H$, apart form the scalar damping term. This fact will be one of the main
ingredients to connect the closed system and the weakly driven open system. For
larger $\mu$, the dynamics is more complicated and also involves the
operators $n_{B_{1}}$ and $n_{B_{2}}$.

Since $H_{\text{eff}}$ is non-Hermitian, the norm of the pure state $\psi(t)$
is not conserved and decays in the course of time, cf.\ Eq.\ (\ref{eq:decay}).
Therefore, at some time $t = \tau$, the condition $|| \psi(t)\rangle \, ||^{2} >
\varepsilon$ is first violated for a given $\varepsilon$, which is here drawn at
random from a box distribution $]0,1]$. At this point in time, a stochastic jump
with one of the Lindblad operators takes place. The new and normalized pure
state reads \cite{Michel2008}
\begin{equation}
| \psi'(t) \rangle = \frac{L_{j}| \psi(t) \rangle}{||L_{j}| \psi(t) \rangle
||} \, ,
\end{equation}
where the specific jump is chosen with probability
\begin{equation} \label{eq:jump_probability}
p_{j} = \frac{\alpha_{j} || L_{j}| \psi(t) \rangle ||^{2}}{\sum_{j}
\alpha_{j} || L_{j} | \psi(t) \rangle ||^{2}} \, .
\end{equation}
After the jump, the procedure continues with the next deterministic time
evolution w.r.t.\ $H_{\text{eff}}$.

This sequence of stochastic jumps and deterministic evolutions leads to a
particular trajectory $|\psi_{T}(t)\rangle$. The time-dependent density matrix
according to the Lindblad equation can be eventually approximated by the average
over different trajectories $T$. Thus, the expectation value reads
\begin{equation}
\langle S_{r}^{z} (t)\rangle \approx \frac{1}{T_{\text{max}}}
\sum_{T=1}^{T_{\text{max}}} \frac{\langle \psi_{T}(t) | S_{r}^{z}
|\psi_{T}(t)\rangle}{|| \, | \psi_{T}(t) \rangle \, ||^{2}} \, ,
\end{equation}
where $T_{\max}$ is a large enough number of trajectories. For $T_{\max} \to
\infty$, the approximation becomes an equality, and the stochastic unraveling
can be used as an in principle exact numerical technique. Moreover, it will
provide the basis for an analytical connection between closed-system and
open-system dynamics.

As already mentioned above, we consider $\rho(0) \propto
1$ as the initial condition of the Lindblad equation, which is realized
in the stochastic unraveling via a random pure state $| \psi(0) \rangle$ of
the form in Eq.\ (\ref{eq:random_state}).


\section{Connection between closed and open systems} \label{sec:connection}

\subsection{Time evolution}

Now, we are ready to discuss our central prediction, which particularly
connects the dynamics in closed and open systems. While this prediction has
been presented in our previous work \cite{Heitmann2022, Heitmann2023} in
detail, we repeat here parts of the derivation, which will be of help to
understand the results presented afterwards.

The derivation is based on the stochastic unraveling of the Lindblad equation.
It is important to recall that we focus on the weak-driving case $\mu \ll 1$.
In this case, the deterministic evolution w.r.t.\ $H_{\text{eff}}$ is unitary
apart from a scalar damping term, cf.\ Eq.\ (\ref{eq:decay}). As a consequence,
when calculating the quantity
\begin{equation}
d_r(t) = \frac{\langle \psi_{T}(t) | S_{r}^{z} |\psi_{T}(t)\rangle}{|| \, |
\psi_{T}(t) \rangle \, ||^{2}} \ ,
\end{equation}
this scalar cancels out. Because the initial condition is a random state of the
form in Eq.\ (\ref{eq:random_state}), the first deterministic evolution is
quite simple. Under unitary time evolution, a random states remains a random
state with a uniform density profile, $d_r(t) = 0$. Therefore, the first
nontrivial event is the subsequent jump.

\begin{figure}[t]
\centering
\includegraphics[width=0.9\columnwidth]{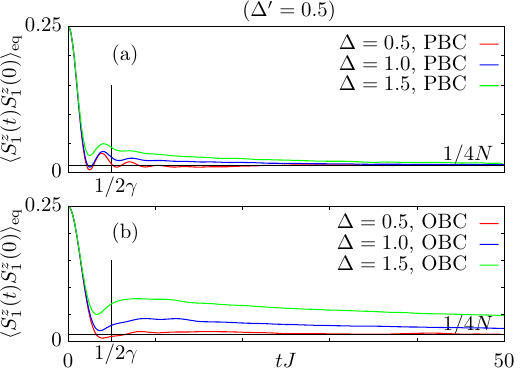}
\caption{Temporal decay of the autocorrelation function
$\langle S_{1}^{z}(t) S_{1}^{z}(0)\rangle_{\text{eq}}$ for (a) periodic
boundary conditions and (b) open boundary conditions, as obtained numerically
for the Hamiltonian $H'$ in Eq.\ (\ref{eq:perturbation1}) for different
$\Delta$ at fixed $\Delta'=0.5$ and $N=20$. The equilibration value $1/(4N)$
and a time scale $1/(2 \gamma)$ with $\gamma/J = 0.1$ are indicated.
}
\label{fig:equal}
\end{figure}

Without loss of generality, let us consider at some time $t = \tau$ a specific
jump with one of the Lindblad operators, e.g., $L_1$. Then, the resulting
pure state reads
\begin{equation}
|\psi' (\tau) \rangle \propto L_{1} \, | \psi(\tau-0^+) \rangle \, ,
\end{equation}
which has exactly the same structure as the pure state in Eq.\
(\ref{eq:L1_state}). Hence, we can employ dynamical quantum typicality and get
\begin{equation}
\frac{d_{r}(t)}{2} \approx \Theta(t - \tau) \, \langle S_{r}^{z}(t - \tau)
S_{B_1}^{z}(0) \rangle_{\text{eq}} \, ,
\end{equation}
where $\Theta(t)$ is the Heavyside function. By the use of the same arguments,
we can obtain analogous relations for the remaining Lindblad operators $L_{j}$,
which then either involve $\langle S_{r}^{z}(t)
S_{B_1}^{z}(0)\rangle_{\text{eq}}$ or $\langle S_{r}^{z}(t)
S_{B_2}^{z}(0)\rangle_{\text{eq}}$. Afterwards, averaging over all four jump
possibilities yields
\begin{eqnarray}
\frac{\bar{d}_{r}(t)}{2} \! &\approx& \! (p_{1} - p_{2}) \, \Theta(t - \tau)
\langle S_{r}^{z}(t - \tau) S_{B_1}^{z}(0) \rangle_{\text{eq}} \nonumber \\
\! &+& \! (p_{3} - p_{4}) \, \Theta(t - \tau) \langle S_{r}^{z}(t - \tau)
S_{B_2}^{z}(0) \rangle_{\text{eq}}
\end{eqnarray}
with jump probabilities $p_{j} = \alpha_{j}/4\gamma$ for a
random state, cf.\ Eq.\ (\ref{eq:jump_probability}). Finally, by inserting the
prefactors $\alpha_j$ from Eqs.\ (\ref{eq:L1}) - (\ref{eq:L4}), we end up with
\begin{eqnarray}
\bar{d}_{r}(t) \! &\approx& \! \mu \, \Theta(t - \tau) \langle
S_{r}^{z}(t - \tau) S_{B_1}^{z}(0) \rangle_{\text{eq}} \nonumber \\[0.1cm]
\! &-& \! \mu \, \Theta(t - \tau) \langle S_{r}^{z}(t - \tau) S_{B_2}^{z}(0)
\rangle_{\text{eq}}
\end{eqnarray}
for the expectation value after the first jump.

\begin{figure}[t]
\centering
\includegraphics[width=0.9\columnwidth]{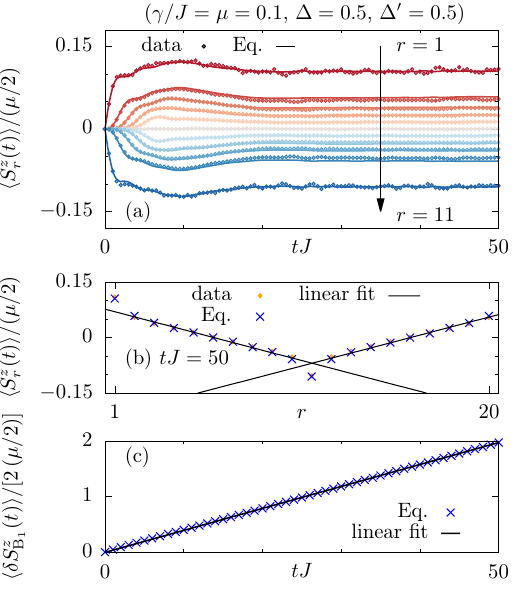}
\caption{Open-system dynamics for the model $H'$ in Eq.\
(\ref{eq:perturbation1}), as obtained numerically for $\Delta = 0.5$, $\Delta'
= 0.5$, $N = 20$, periodic boundary conditions, small coupling $\gamma/J = 0.1$,
and weak driving $\mu = 0.1$. Exact results from the full stochastic
unraveling (data) are compared to the prediction (Eq.), which is based on
spatio-temporal correlation functions in the closed system. (a) Time evolution
of the local magnetization $\langle S_{r}^{z}(t)\rangle$ for different sites
$r$. (b) Site dependence of the steady state at $tJ=50$. (c) Magnetization
injected by the first bath as a function of time.}
\label{fig:PBC_D0_5_DTwo0_5}
\end{figure}

To proceed, a natural idea is to reuse the same line of reasoning for the
second and all subsequent jumps. After the first jump, however, the pure state
is different, since it has an inhomogeneous density profile with magnetization
concentrated at the site of the bath contact. Thus, one has to wait until the
injected magnetization has spread over a piece of the system. Clearly, such a
waiting time requires a small enough value of the system-bath coupling $\gamma$,
as illustrated in Fig.\ \ref{fig:equal}. This kind of equilibration is
a central ingredient, and its impact will be scrutinized for specific models
later. Assuming equilibration, we can iterate the arguments and
obtain a superposition of the form
\begin{eqnarray} \label{eq:superposition}
\frac{\bar{d}_r(t)}{2 \mu} \approx \sum_j && A_j \, \Theta(t -
\tau_j) \, \Big [ \langle S_r^z(t - \tau_j) S_{B_1}^z(0) \rangle_\text{eq}
\nonumber \\
&& - \langle S_r^z(t - \tau_j) S_{B_2}^z(0) \rangle_\text{eq} \Big ] \, ,
\end{eqnarray}
where the amplitudes $A_j$ can be calculated from the jump probabilities in Eq.\
(\ref{eq:jump_probability}) and result as \cite{Heitmann2023}
\begin{equation}
A_j = \frac{a_j - \bar{d}_{B_1}(\tau_j - 0^+)}{\mu}
\label{eq:amplitudes}
\end{equation}
with
\begin{equation}
a_j = \frac{\mu - 2 \, \bar{d}_{B_1}(\tau_j - 0^+)}{2 - 4 \mu \,
\bar{d}_{B_1}(\tau_j - 0^+)} \, .
\end{equation}
If $\bar{d}_{B_1}(\tau_j - 0^+) \to 0$, $A_j \to 1/2$.

\begin{figure}[t]
\centering
\includegraphics[width=0.9\columnwidth]{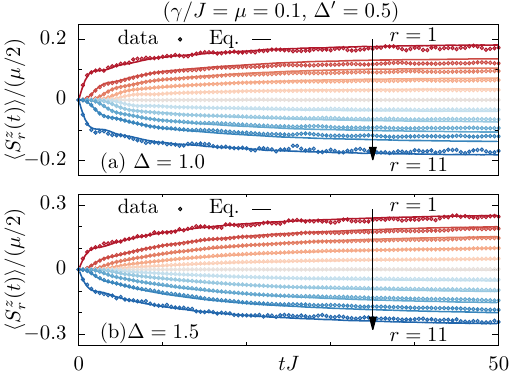}
\caption{Analogous data as the one in Fig.\ \ref{fig:PBC_D0_5_DTwo0_5} (a), but
now for (a) $\Delta=1.0$ and (b) $\Delta=1.5$.}
\label{fig:PBC_D_1_0_and_D1_5}
\end{figure}

Because the expression in Eq.\ (\ref{eq:superposition}) applies to a single
sequence of jump times, ($\tau_1, \tau_2, \ldots$), the final prediction is
obtained by the average
\begin{equation} \label{eq:final_prediction}
\langle S_{r}^{z}(t)\rangle \approx \frac{1}{T_{\text{max}}}
\sum_{T=1}^{T_{\text{max}}} \bar{d}_{r,T}(t)
\end{equation}
over trajectories with different jump times. Due to the scalar damping
term in Eq.\ (\ref{eq:decay}), these jump times are given by
\begin{equation} \label{eq:tau}
\tau_{j+1} = \tau_j - \ln \frac{\varepsilon_{j+1}}{2 \gamma} \, ,
\end{equation}
where $\varepsilon_{j+1}$ is drawn at random from a box distribution
$]0,1]$ again.

In principle, the prediction in Eqs.\ (\ref{eq:superposition}) and
(\ref{eq:final_prediction}) can be calculated analytically for a specific
model. However, the closed-system correlation functions $\langle
S_{r}^{z}(t) S_{B_1}^{z}(0) \rangle_{\text{eq}}$ and $\langle S_{r}^{z}(t)
S_{B_2}^{z}(0) \rangle_{\text{eq}}$ are often available only numerically, such
that the prediction has to be calculated numerically as well.

\begin{figure}[t]
\centering
\includegraphics[width=0.9\columnwidth]{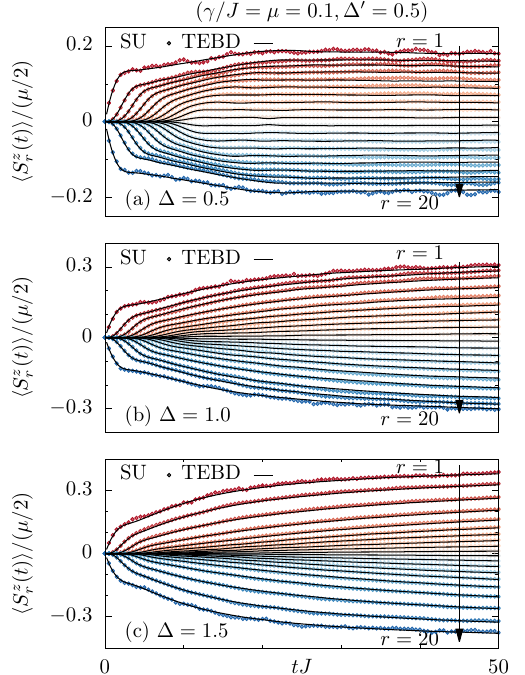}
\caption{Comparison of the solution of the Lindblad equation, as obtained
numerically from stochastic unraveling (SU) and time-evolving block decimation
(TEBD), for the Hamiltonian $H'$ in Eq.\ (\ref{eq:perturbation1}) with (a)
$\Delta = 0.5$, (b) $\Delta=1.0$, and (c) $\Delta=1.5$ as well as $N=20$,
open boundary conditions, $\gamma/J = 0.1$, and $\mu = 0.1$.}
\label{fig:OBC_TEBD_vs_SU_L20}
\end{figure}

\subsection{Injected magnetization}

While Eqs.\ (\ref{eq:superposition}) and (\ref{eq:final_prediction}) allow
to predict the dynamics of magnetization at finite times and in the limit of
long times, a similar expression can derived for the respective currents in
the steady state. To this end, let us consider the magnetization injected by
the first bath, which can be predicted as \cite{Heitmann2023}
\begin{equation}
\langle\delta S^{z}_{B_1}(t) \rangle \approx \frac{1}{T_{\text{max}}}
\sum_{T=1}^{T_{\text{max}}} \delta \bar{d}_{B_{1},T}(t)
\end{equation}
with $\delta$ being just a notation for ``injected'' and
\begin{equation}
\frac{\delta \bar{d}_{B_{1},T}(t) }{2\mu} \approx \sum_{j} A_{j} \,
\Theta(t-\tau_{j}) \, \langle[S_{B_{1}}^{z}(0)]^{2} \rangle \, ,
\end{equation}
which is slightly simpler than Eq.\ (\ref{eq:superposition}). Since in the
steady state all currents are the same,
\begin{equation}
\langle j_{r} \rangle = \langle j_{r'}\rangle \, , \quad B_{1} \leq r,r' \leq
B_{2} \, ,
\end{equation}
it is sufficient to know $\langle j_{B_1} \rangle$, which can expressed as
\begin{equation}
\langle j_{B_{1}}\rangle = \frac{\text{d}}{\text{d} t} \frac{\langle
\delta S_{B_{1}}^{z}(t) \rangle}{f} \, .
\end{equation}
Here, $f = 2$ for periodic boundary conditions (flow to the right and left of
the bath) and $f = 1$ for open boundary conditions (flow only to the right of
the bath).

With the knowledge of the steady-state current, it is also possible to predict
the diffusion constant via \cite{Bertini2021}
\begin{eqnarray} \label{eq:Fourier}
D = - \frac{\langle j_{r} \rangle}{\langle S_{r+1}^{z} \rangle - \langle
S_{r}^{z} \rangle}
\end{eqnarray}
for some site $r$ in the bulk.


\section{Results} \label{sec:results}

\subsection{Next-to-nearest-neighbor interactions and periodic boundary
conditions}

Finally, we turn to our numerical simulations, where the central goal is to
analyze the quality of the prediction in Eqs.\ (\ref{eq:superposition}) and
(\ref{eq:final_prediction}) for various situations. To start with, we
investigate the spin-$1/2$ XXZ chain with interactions between next-to-nearest
neighbors, Eq.\ (\ref{eq:perturbation1}), and the case of periodic boundary
conditions, Fig.\ \ref{fig:sketch1} (a). Afterwards, we additionally study other
perturbations and the case of open boundary conditions with different
bath-coupling geometries.

\begin{figure}[t]
\centering
\includegraphics[width=0.9\columnwidth]{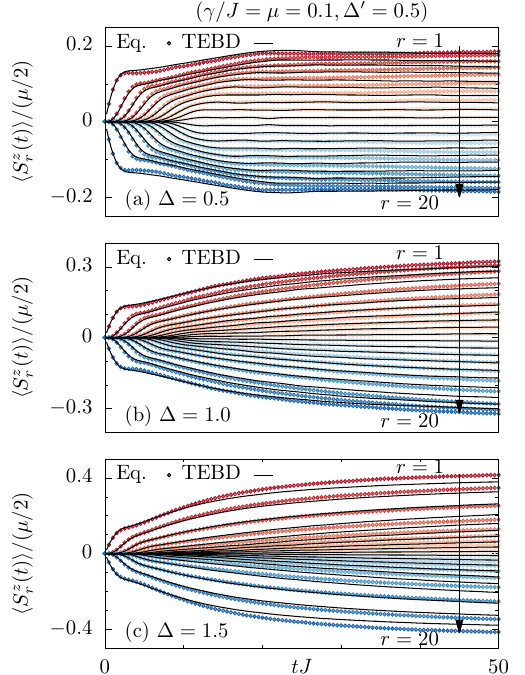}
\caption{Similar comparison as the one in Fig.\ \ref{fig:OBC_TEBD_vs_SU_L20},
but now, instead of stochastic unraveling, between the prediction (Eq.) and
time-evolving block decimation (TEBD).}
\label{fig:OBC_TEBD_vs_Eq_L20}
\end{figure}

Because a main ingredient of the prediction has been the equilibration of
the injected magnetization, we first focus on this assumption. To this
end, we numerically calculate in Fig.\ \ref{fig:equal} (a) the equal-site
correlation function $\langle S_{r}^{z}(t) S_{r}^{z}(0) \rangle_{\text{eq}}$ for
$\Delta' = 0.5$, $N = 20$, and an arbitrary site $r$ due to periodic
boundary conditions. Apparently, for all $\Delta$ depicted, this function starts
with the initial value $1/4$, decays substantially on a time scale $t_R J
\approx 5$, and then approaches the equilibration value $1/(4 N)$ in the limit
of long times. By comparing $t_R$ to $1/(2 \gamma)$ from Eq.\ (\ref{eq:tau}),
we identify $\gamma/J = 0.1$ as a reasonable choice for the system-bath
coupling, which we fix from now on for a fair comparison, together with the
choice $\mu = 0.1$ to ensure weak driving.

\begin{figure}[t]
\centering
\includegraphics[width=0.9\columnwidth]{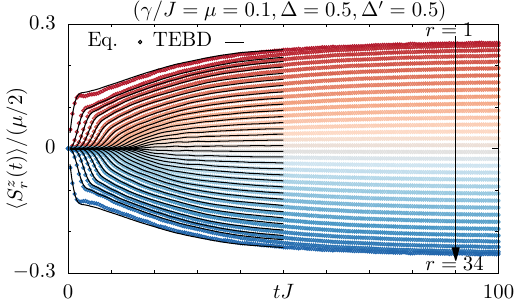}
\caption{Similar comparison as the one in Fig.\ \ref{fig:OBC_TEBD_vs_Eq_L20}
(a), but now for system size $N=34 \gg 20$, where stochastic unraveling is not
possible any more.}
\label{fig:OBC_L34}
\end{figure}

For the value $\Delta = 0.5$, we depict in Fig.\
\ref{fig:PBC_D0_5_DTwo0_5} (a) the time evolution of magnetization in the open
system. Here, the prediction is carried out for $\approx 10,000$ different
sequences of jump times, which already yields smooth curves. The full stochastic
unraveling, without any assumption, has been evaluated on clusters and
turns out to require as many as $\approx 200,000$ (or more) sequences for a
comparable smoothness \cite{note2}. Despite residual statistical
fluctuations, the agreement is almost perfect and clearly visible from short
until long times. This convincing agreement can also be seen for the
steady-state profile in Fig.\ \ref{fig:PBC_D0_5_DTwo0_5} (b). Compared to
previous results for integrable systems \cite{Heitmann2023}, the degree of
agreement turns out to be similar. This finding indicates that nonintegrability
is not required for the prediction to hold, which is consistent with the fact
that such an assumption does not enter the derivation.

In Fig.\ \ref{fig:PBC_D0_5_DTwo0_5} (c), we also show the injected
magnetization, which grows linearly with time. From the slope, and the slope of
the steady-state density profile in Fig.\ \ref{fig:PBC_D0_5_DTwo0_5} (b), we can
extract a diffusion constant via Eq.\ (\ref{eq:Fourier}). The value $D/J
\approx 2.9$ agrees well with other values in the literature, such as $D/J
\approx 3.1$ in the closed system \cite{Wang2023, Richter2019}, and serves as a
further sanity check of our approach.

To ensure that our results do not depend on the specific choice of parameters,
we redo in Fig.\ \ref{fig:PBC_D_1_0_and_D1_5} the calculation for other $\Delta
\neq 0.5$. The overall agreement is apparently the same.

\subsection{Open boundary conditions and different bath-coupling geometries}

\begin{figure}[t]
\centering
\includegraphics[width=0.9\columnwidth]{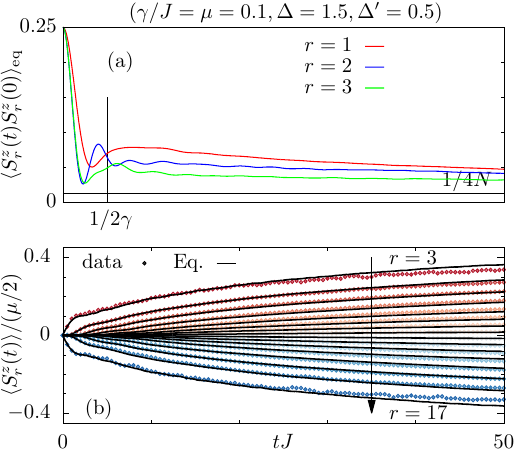}
\caption{On a bath coupling close to the edges. (a) Temporal decay
of the autocorrelation function $\langle S_{r}^{z}(t)
S_{r}^{z}(0)\rangle_{\text{eq}}$ for different sites $r$, as obtained
numerically for the model $H'$ in Eq.\ (\ref{eq:perturbation1}) with
$\Delta = 1.5$, $\Delta'=0.5$, $N=20$, and open boundary conditions. (b)
Open-system dynamics for a bath coupling at sites $B_1 = 3$ and $B_2 =
17$. In comparison to the data in Fig.\ \ref{fig:OBC_TEBD_vs_Eq_L20} (c), the
agreement between the prediction (Eq.) and stochastic unraveling (data) is
better.}
\label{fig:OBC_D1_5_r3}
\end{figure}

Now, we move forward to the case of open boundary conditions with a standard
bath coupling at the edges, as sketched in Fig.\ \ref{fig:sketch1} (b). For
such a situation, it is also possible to obtain the solution of the Lindblad
equation by TEBD \cite{Schollwoeck2011,
Verstraete2008}, as a state-of-the-art technique in this context. For the same
Hamiltonian and parameters as before, we depict in Fig.\
\ref{fig:OBC_TEBD_vs_SU_L20} the numerical result from TEBD and additionally
compare to the exact stochastic-unraveling procedure. Even though statistical
fluctuations are again visible, both approaches coincide for all times and
$\Delta$ depicted. This agreement particularly confirms the correctness of our
numerics.

When we compare TEBD to the actual prediction in Fig.\
\ref{fig:OBC_TEBD_vs_Eq_L20}, the agreement is less convincing for long times
and becomes worse for larger values of $\Delta$. To understand the origin of
the disagreement, we test the assumption of equilibration, by calculating in
Fig.\ \ref{fig:equal} (b) the equal-site correlation function $\langle
S_{r}^{z}(t) S_{r}^{z}(0) \rangle_{\text{eq}}$ at the left-edge site $r=1$. In
contrast to periodic boundary conditions, this function decays slower and
develops long-time tails for large $\Delta$, which can be traced back to
nondecaying edge modes occurring for $\Delta > 1$ \cite{Fendley2016, Kemp2017}.
As the occurrence of long-time tails in Fig.\ \ref{fig:equal} (b) seems to
correlate with the degree of disagreement in Fig.\
\ref{fig:OBC_TEBD_vs_Eq_L20}, we can identify the breakdown of the assumption
as the origin.

For $\Delta = 0.5$, where the assumption is fulfilled best, we increase next
the system size to $N=34 \gg 20$. For such a system size, the Hilbert-space
dimension becomes huge and stochastic unraveling is not feasible any more,
due to the many trajectories required. However, the prediction can still be
carried out, since the two correlation functions $\langle S_{r}^{z}(t)
S_{B_1}^{z}(0)\rangle_{\text{eq}}$ and $\langle S_{r}^{z}(t)
S_{B_2}^{z}(0)\rangle_{\text{eq}}$ need to be calculated only once. In
particular, this calculation is possible by the use of dynamical quantum
typicality and supercomputers. In Fig.\ \ref{fig:OBC_L34}, we depict the
corresponding prediction and compare to the TEBD solution, which up to times $t
J \approx 50$ does not depend on the bond dimension used. (A convergence
analysis of TEBD can be found in the appendix.) The convincing agreement
supports that our prediction is a useful alternative for large system sizes,
which are usually only accessible by matrix product states.

Unfortunately, our assumption is not always satisfied for open boundary
conditions, as discussed above. Hence, we explore possibilities to
circumvent this problem. To this end, we consider the slightly different
bath-coupling geometry in Fig.\ \ref{fig:sketch1} (c), where the Lindblad
operators are not located exactly at the edges, but close to them. As depicted
in Fig.\ \ref{fig:OBC_D1_5_r3} (a), the equal-site correlation function $\langle
S_{r}^{z}(t) S_{r}^{z}(0) \rangle_{\text{eq}}$ tends to decay stronger when the
site $r$ is moved away from the left edge $r=1$, indicating a higher degree of
equilibration. And indeed, for the so far worst case $\Delta = 1.5$, the
prediction for the open-system dynamics in Fig.\ \ref{fig:OBC_D1_5_r3} (b)
agrees substantially better with the full stochastic unraveling. This
observation supports the usefulness of other bath-coupling geometries, which
have attracted less attention yet.

\subsection{Staggered field}

\begin{figure}[t]
\centering
\includegraphics[width=0.9\columnwidth]{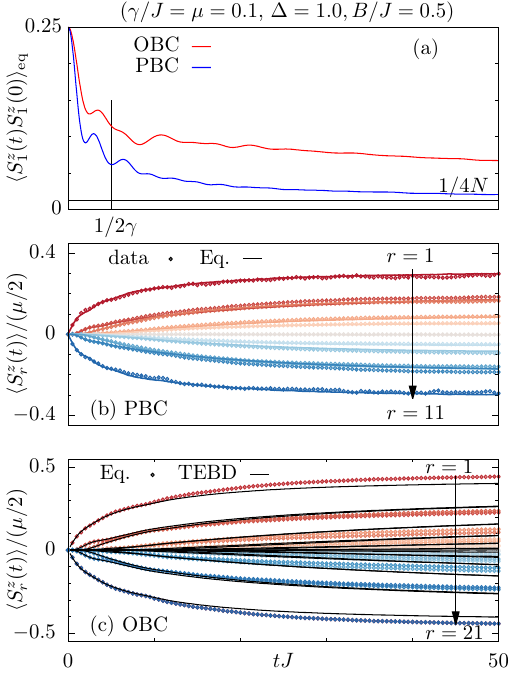}
\caption{Another model: staggered field. (a) Temporal decay of the autocorrelation function
$\langle S_{1}^{z}(t) S_{1}^{z}(0)\rangle_{\text{eq}}$ for
closed and open boundary conditions, as obtained numerically for the Hamiltonian
$H''$ in Eq.\ (\ref{eq:perturbation2}) with $\Delta = 1.0$, $B/J = 0.5$,
and $N=20$. (b) and (c) Open-system dynamics for the respective boundary
conditions as well as $\gamma/J = 0.1$ and $\mu = 0.1$. The prediction
(Eq.) is compared to either stochastic unraveling (data) or time-evolving block
decimation (TEBD).}
\label{fig:Zeeman}
\end{figure}

Eventually, we also study other perturbations and turn to the spin-$1/2$ XXZ
chain with a staggered field, Eq.\ (\ref{eq:perturbation2}), where we focus on
$\Delta = 1.0$, $B/J = 0.5$, and $N = 20$. In Fig.\ \ref{fig:Zeeman}, we
summarize our numerical results. Apparently, the situation is overall similar.
The equal-site correlation function $\langle S_{r}^{z}(t) S_{r}^{z}(0)
\rangle_{\text{eq}}$ in Fig.\ \ref{fig:Zeeman} (a) behaves differently
for periodic and open boundary conditions. Consistently, the
prediction for the open-system dynamics agrees well with numerics for the
periodic-boundaries case in Fig.\ \ref{fig:Zeeman} (b), while deviations
are visible for the open-boundaries case in Fig.\ \ref{fig:Zeeman} (c). We have
checked that the situation remains the same for other parameters of
$\Delta$, even though not explicitly shown here.

\begin{figure}[t]
\centering
\includegraphics[width=0.9\columnwidth]{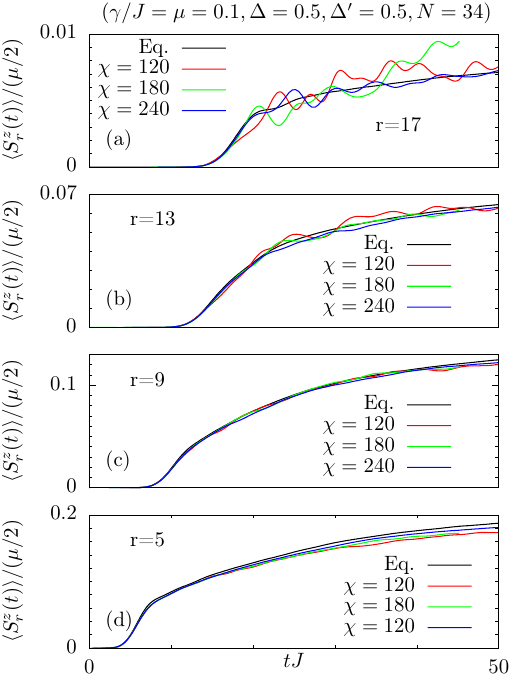}
\caption{Convergence analysis. The data in Fig.\ \ref{fig:OBC_L34} is depicted
again for different sites: (a) $r = 17$, (b) $r = 13$, (c) $r = 9$, and (d) $r
= 5$. But now, the prediction (Eq.) is compared to data from time-evolving
block decimation (TEBD) for various bond dimensions $\chi$. While $r = 5$ is
close to the edge, $r = 17$ lies in the bulk.}
\label{fig:convergence_L34_TEBD}
\end{figure}


\section{Conclusion} \label{sec:conslusion}

To summarize, we have studied the Lindblad equation as a central approach to
boundary-driven magnetization transport in spin-$1/2$ chains. Our main
motivation has been to understand to what extent the dynamics
in the open
system, at finite times and in the limit of long times, can be predicted on the
basis of the dynamics in the closed system. To this end, we have followed the
idea of a previous work \cite{Heitmann2023}, which has suggested a prediction
in terms of spatio-temporal correlation functions, Eqs.\
(\ref{eq:superposition}) and (\ref{eq:final_prediction}), given the case of
weak driving and small system-bath coupling. While this work was focused on
integrable systems and periodic boundary conditions, we have substantially
extended the analysis in the current work by going in three different
directions: We have (i) considered nonintegrable systems, (ii) included open
boundary conditions and other bath-coupling geometries, and (iii) provided a
comparison with time-evolving block decimation.

We have found that nonintegrability plays a minor role, since the quality of the
prediction is comparable to the one for integrable systems. This observation is
consistent with the fact that nonintegrability does not enter as an assumption
in the derivation. In contrast, the choice of the specific boundary conditions
has turned out to be of relevance. For periodic boundary conditions, on the one
hand, prediction and numerical simulations have agreed convincingly, for all
models and parameters considered here. For open boundary conditions, on the
other hand, we have observed some disagreement in particular cases, which we
have traced back to slowly decaying edge modes and thus a breakdown of the
equilibration assumption underlying the prediction. In this context, it is
important to note that the validity of the assumption can be checked in advance
and does not require the comparison to other methods. To circumvent such edge
modes, we have also explored other bath-coupling geometries, where the Lindblad
operators are not acting exactly at the boundary sites, but still close to
them.

For parameters, where the assumption is well fulfilled also for open boundary
conditions, we have demonstrated that the prediction yields an accurate
description and can be carried out for comparatively large system sizes, which
are usually accessible by matrix-product-states methods only. From a less
practical but more physical perspective, we have thus shown a kind
of one-to-one correspondence between the time evolution in open and
closed systems, at least for the models considered by us.

Promising future directions of future research include  quasi-1D lattices,
finite temperatures, energy transport, fermionic models, and disorder.
Another interesting avenue is to explore in which cases the
slowly decaying edge modes of the closed system can be enhanced in the open
system \cite{Vasiloiu2018}.

\begin{figure}[t]
\centering
\includegraphics[width=0.9\columnwidth]{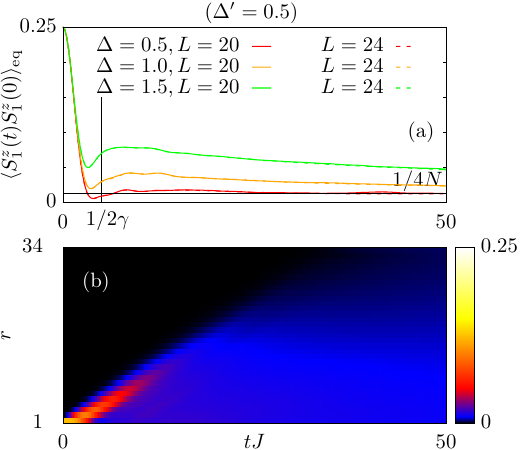}
\caption{(a) Dynamics of the autocorrelation function $\langle
S_{1}^{z}(t) S_{1}^{z}(0)\rangle_{\text{eq}}$ for open boundary conditions, as
depicted in Fig.\ \ref{fig:equal} (b) for different $\Delta$, but now for two
system sizes $N = 20$ and $N = 24$. (b) Time-space density plot of
the correlation functions $\langle S_{r}^{z}(t)
S_{1}^{z}(0)\rangle_{\text{eq}}$ for $N = 34$ and $\Delta = 0.5$, which are
used for the prediction in Fig.\ \ref{fig:OBC_L34}.}
\label{fig:edege_correlations}
\end{figure}


\subsection*{Acknowledgments}

Our work has been funded by the Deutsche Forschungsgemeinschaft (DFG), projects
397107022 (GE 1657/3-2), 397300368 (MI 1772/4-2), and 397067869 (STE 2243/3-2),
within DFG Research Unit FOR 2692, grant no.\ 355031190. J.\ R.\ acknowledges
funding from the European Union’s Horizon Europe research and innovation
programme, Marie Sklodowska-Curie grant no.\ 101060162, and the Packard
Foundation through a Packard Fellowship in Science and Engineering. We
gratefully acknowledge the Gauss Centre for Supercomputing e.\ V.\ for funding
this project by providing computing time on the GCS Supercomputer JUWELS6 at
J\"ulich Supercomputing Centre (JSC). S.\ N.\
was supported by QuantERA grants QuSiED and T-NiSQ,
by MVZI, QuantERA II JTC 2021. TEBD calculations were
performed on the supercomputer Vega at the Institute of Infor-
mation Science (IZUM) in Maribor, Slovenia.

\appendix*


\section{Convergence of the TEBD method}

In the main text, we have shown in Fig.\ \ref{fig:OBC_L34} numerical data from
TEBD and stated that, up to times $t J \approx 50$, it does not depend on
the bond dimension $\chi$ used. To further substantiate this statement, we
depict in Fig.\ \ref{fig:convergence_L34_TEBD} the same data for different
$\chi$ and various sites, close to the edges and in the bulk. While the data is
particularly well converged close to the edges, some oscillations can be seen
in the bulk, where the time evolution is kind of close to unitary.

\section{Correlation functions for open boundary conditions}

In Fig.\ \ref{fig:equal} (b), we have shown the equal-site
correlation function $\langle S_{r}^{z}(t) S_{r}^{z}(0)\rangle_{\text{eq}}$ for
the open-boundaries case and different $\Delta$, where we have focused on a
single system size $N = 20$. To demonstrate that the temporal decay does not
depend significantly on system size, we additionally depict in Fig.\
\ref{fig:edege_correlations} (a) numerical data for $N = 24$.

For completeness, Fig.\ \ref{fig:edege_correlations} (b) shows the full
time-space dependence of the correlation functions $\langle S_{r}^{z}(t)
S_{r'}^{z}(0)\rangle_{\text{eq}}$ for $N = 34$ and $\Delta = 0.5$, which have
been used for the prediction in Fig.\ \ref{fig:OBC_L34}.



\begin{thebibliography}{65}%
\makeatletter
\providecommand \@ifxundefined [1]{%
 \@ifx{#1\undefined}
}%
\providecommand \@ifnum [1]{%
 \ifnum #1\expandafter \@firstoftwo
 \else \expandafter \@secondoftwo
 \fi
}%
\providecommand \@ifx [1]{%
 \ifx #1\expandafter \@firstoftwo
 \else \expandafter \@secondoftwo
 \fi
}%
\providecommand \natexlab [1]{#1}%
\providecommand \enquote  [1]{``#1''}%
\providecommand \bibnamefont  [1]{#1}%
\providecommand \bibfnamefont [1]{#1}%
\providecommand \citenamefont [1]{#1}%
\providecommand \href@noop [0]{\@secondoftwo}%
\providecommand \href [0]{\begingroup \@sanitize@url \@href}%
\providecommand \@href[1]{\@@startlink{#1}\@@href}%
\providecommand \@@href[1]{\endgroup#1\@@endlink}%
\providecommand \@sanitize@url [0]{\catcode `\\12\catcode `\$12\catcode
  `\&12\catcode `\#12\catcode `\^12\catcode `\_12\catcode `\%12\relax}%
\providecommand \@@startlink[1]{}%
\providecommand \@@endlink[0]{}%
\providecommand \url  [0]{\begingroup\@sanitize@url \@url }%
\providecommand \@url [1]{\endgroup\@href {#1}{\urlprefix }}%
\providecommand \urlprefix  [0]{URL }%
\providecommand \Eprint [0]{\href }%
\providecommand \doibase [0]{http://dx.doi.org/}%
\providecommand \selectlanguage [0]{\@gobble}%
\providecommand \bibinfo  [0]{\@secondoftwo}%
\providecommand \bibfield  [0]{\@secondoftwo}%
\providecommand \translation [1]{[#1]}%
\providecommand \BibitemOpen [0]{}%
\providecommand \bibitemStop [0]{}%
\providecommand \bibitemNoStop [0]{.\EOS\space}%
\providecommand \EOS [0]{\spacefactor3000\relax}%
\providecommand \BibitemShut  [1]{\csname bibitem#1\endcsname}%
\let\auto@bib@innerbib\@empty
\bibitem [{\citenamefont {Bloch}\ \emph {et~al.}(2008)\citenamefont {Bloch},
  \citenamefont {Dalibard},\ and\ \citenamefont {Zwerger}}]{Bloch2008}%
  \BibitemOpen
  \bibfield  {author} {\bibinfo {author} {\bibfnamefont {I.}~\bibnamefont
  {Bloch}}, \bibinfo {author} {\bibfnamefont {J.}~\bibnamefont {Dalibard}}, \
  and\ \bibinfo {author} {\bibfnamefont {W.}~\bibnamefont {Zwerger}},\
  }\bibfield  {title} {\emph {\bibinfo {title} {{Many-body physics with
  ultracold gases}},\ }}\href {https://doi.org/10.1103/RevModPhys.80.885}
  {\bibfield  {journal} {\bibinfo  {journal} {Rev. Mod. Phys.}\ }\textbf
  {\bibinfo {volume} {80}},\ \bibinfo {pages} {885} (\bibinfo {year}
  {2008})}\BibitemShut {NoStop}%
\bibitem [{\citenamefont {Abanin}\ \emph {et~al.}(2019)\citenamefont {Abanin},
  \citenamefont {Altman}, \citenamefont {Bloch},\ and\ \citenamefont
  {Serbyn}}]{Abanin2019}%
  \BibitemOpen
  \bibfield  {author} {\bibinfo {author} {\bibfnamefont {D.~A.}\ \bibnamefont
  {Abanin}}, \bibinfo {author} {\bibfnamefont {E.}~\bibnamefont {Altman}},
  \bibinfo {author} {\bibfnamefont {I.}~\bibnamefont {Bloch}}, \ and\ \bibinfo
  {author} {\bibfnamefont {M.}~\bibnamefont {Serbyn}},\ }\bibfield  {title}
  {\emph {\bibinfo {title} {{Colloquium: Many-body localization,
  thermalization, and entanglement}},\ }}\href
  {https://doi.org/10.1103/RevModPhys.91.021001} {\bibfield  {journal}
  {\bibinfo  {journal} {Rev. Mod. Phys.}\ }\textbf {\bibinfo {volume} {91}},\
  \bibinfo {pages} {021001} (\bibinfo {year} {2019})}\BibitemShut {NoStop}%
\bibitem [{\citenamefont {Polkovnikov}\ \emph {et~al.}(2011)\citenamefont
  {Polkovnikov}, \citenamefont {Sengupta}, \citenamefont {Silva},\ and\
  \citenamefont {Vengalattore}}]{Polkovnikov2011}%
  \BibitemOpen
  \bibfield  {author} {\bibinfo {author} {\bibfnamefont {A.}~\bibnamefont
  {Polkovnikov}}, \bibinfo {author} {\bibfnamefont {K.}~\bibnamefont
  {Sengupta}}, \bibinfo {author} {\bibfnamefont {A.}~\bibnamefont {Silva}}, \
  and\ \bibinfo {author} {\bibfnamefont {M.}~\bibnamefont {Vengalattore}},\
  }\bibfield  {title} {\emph {\bibinfo {title} {{Colloquium: Nonequilibrium
  dynamics of closed interacting quantum systems}},\ }}\href
  {https://doi.org/10.1103/RevModPhys.83.863} {\bibfield  {journal} {\bibinfo
  {journal} {Rev. Mod. Phys.}\ }\textbf {\bibinfo {volume} {83}},\ \bibinfo
  {pages} {863} (\bibinfo {year} {2011})}\BibitemShut {NoStop}%
\bibitem [{\citenamefont {Eisert}\ \emph {et~al.}(2015)\citenamefont {Eisert},
  \citenamefont {Friesdorf},\ and\ \citenamefont {Gogolin}}]{Eisert2015}%
  \BibitemOpen
  \bibfield  {author} {\bibinfo {author} {\bibfnamefont {J.}~\bibnamefont
  {Eisert}}, \bibinfo {author} {\bibfnamefont {M.}~\bibnamefont {Friesdorf}}, \
  and\ \bibinfo {author} {\bibfnamefont {C.}~\bibnamefont {Gogolin}},\
  }\bibfield  {title} {\emph {\bibinfo {title} {{Quantum many-body systems out
  of equilibrium}},\ }}\href {\doibase 10.1038/nphys3215} {\bibfield  {journal}
  {\bibinfo  {journal} {Nat. Phys.}\ }\textbf {\bibinfo {volume} {11}},\
  \bibinfo {pages} {124} (\bibinfo {year} {2015})}\BibitemShut {NoStop}%
\bibitem [{\citenamefont {D'Alessio}\ \emph {et~al.}(2016)\citenamefont
  {D'Alessio}, \citenamefont {Kafri}, \citenamefont {Polkovnikov},\ and\
  \citenamefont {Rigol}}]{Dalessio2016}%
  \BibitemOpen
  \bibfield  {author} {\bibinfo {author} {\bibfnamefont {L.}~\bibnamefont
  {D'Alessio}}, \bibinfo {author} {\bibfnamefont {Y.}~\bibnamefont {Kafri}},
  \bibinfo {author} {\bibfnamefont {A.}~\bibnamefont {Polkovnikov}}, \ and\
  \bibinfo {author} {\bibfnamefont {M.}~\bibnamefont {Rigol}},\ }\bibfield
  {title} {\emph {\bibinfo {title} {{From quantum chaos and eigenstate
  thermalization to statistical mechanics and thermodynamics}},\ }}\href
  {\doibase 10.1080/00018732.2016.1198134} {\bibfield  {journal} {\bibinfo
  {journal} {Adv. Phys.}\ }\textbf {\bibinfo {volume} {65}},\ \bibinfo {pages}
  {239} (\bibinfo {year} {2016})}\BibitemShut {NoStop}%
\bibitem [{\citenamefont {Deutsch}(1991)}]{Deutsch1991}%
  \BibitemOpen
  \bibfield  {author} {\bibinfo {author} {\bibfnamefont {J.~M.}\ \bibnamefont
  {Deutsch}},\ }\bibfield  {title} {\emph {\bibinfo {title} {Quantum
  statistical mechanics in a closed system},\ }}\href
  {https://doi.org/10.1103/PhysRevA.43.2046} {\bibfield  {journal} {\bibinfo
  {journal} {Phys. Rev. A}\ }\textbf {\bibinfo {volume} {43}},\ \bibinfo
  {pages} {2046} (\bibinfo {year} {1991})}\BibitemShut {NoStop}%
\bibitem [{\citenamefont {Srednicki}(1994)}]{Srednicki1994}%
  \BibitemOpen
  \bibfield  {author} {\bibinfo {author} {\bibfnamefont {M.}~\bibnamefont
  {Srednicki}},\ }\bibfield  {title} {\emph {\bibinfo {title} {Chaos and
  quantum thermalization},\ }}\href {\doibase 10.1103/PhysRevE.50.888}
  {\bibfield  {journal} {\bibinfo  {journal} {Phys. Rev. E}\ }\textbf {\bibinfo
  {volume} {50}},\ \bibinfo {pages} {888} (\bibinfo {year} {1994})}\BibitemShut
  {NoStop}%
\bibitem [{\citenamefont {Rigol}\ \emph {et~al.}(2008)\citenamefont {Rigol},
  \citenamefont {Dunjko},\ and\ \citenamefont {Olshanii}}]{Rigol2008}%
  \BibitemOpen
  \bibfield  {author} {\bibinfo {author} {\bibfnamefont {M.}~\bibnamefont
  {Rigol}}, \bibinfo {author} {\bibfnamefont {V.}~\bibnamefont {Dunjko}}, \
  and\ \bibinfo {author} {\bibfnamefont {M.}~\bibnamefont {Olshanii}},\
  }\bibfield  {title} {\emph {\bibinfo {title} {Thermalization and its
  mechanism for generic isolated quantum systems},\ }}\href
  {https://doi.org/10.1038/nature06838} {\bibfield  {journal} {\bibinfo
  {journal} {Nature}\ }\textbf {\bibinfo {volume} {452}},\ \bibinfo {pages}
  {854} (\bibinfo {year} {2008})}\BibitemShut {NoStop}%
\bibitem [{\citenamefont {Bartsch}\ and\ \citenamefont
  {Gemmer}(2009)}]{Bartsch2009}%
  \BibitemOpen
  \bibfield  {author} {\bibinfo {author} {\bibfnamefont {C.}~\bibnamefont
  {Bartsch}}\ and\ \bibinfo {author} {\bibfnamefont {J.}~\bibnamefont
  {Gemmer}},\ }\bibfield  {title} {\emph {\bibinfo {title} {{Dynamical
  typicality of quantum expectation values}},\ }}\href
  {https://doi.org/10.1103/PhysRevLett.102.110403} {\bibfield  {journal}
  {\bibinfo  {journal} {Phys. Rev. Lett.}\ }\textbf {\bibinfo {volume} {102}},\
  \bibinfo {pages} {110403} (\bibinfo {year} {2009})}\BibitemShut {NoStop}%
\bibitem [{\citenamefont {Gemmer}\ \emph {et~al.}(2004)\citenamefont {Gemmer},
  \citenamefont {Michel},\ and\ \citenamefont {Mahler}}]{Gemmer2004}%
  \BibitemOpen
  \bibfield  {author} {\bibinfo {author} {\bibfnamefont {J.}~\bibnamefont
  {Gemmer}}, \bibinfo {author} {\bibfnamefont {M.}~\bibnamefont {Michel}}, \
  and\ \bibinfo {author} {\bibfnamefont {G.}~\bibnamefont {Mahler}},\ }\href
  {\doibase 10.1007/b98082} {\emph {\bibinfo {title} {{Quantum
  thermodynamics}}}},\ \bibinfo {series} {Lect. Notes Phys.}, Vol.\ \bibinfo
  {volume} {657}\ (\bibinfo  {publisher} {Springer},\ \bibinfo {year}
  {2004})\BibitemShut {NoStop}%
\bibitem [{\citenamefont {Goldstein}\ \emph {et~al.}(2006)\citenamefont
  {Goldstein}, \citenamefont {Lebowitz}, \citenamefont {Tumulka},\ and\
  \citenamefont {Zangh{\`{i}}}}]{Goldstein2006}%
  \BibitemOpen
  \bibfield  {author} {\bibinfo {author} {\bibfnamefont {S.}~\bibnamefont
  {Goldstein}}, \bibinfo {author} {\bibfnamefont {J.~L.}\ \bibnamefont
  {Lebowitz}}, \bibinfo {author} {\bibfnamefont {R.}~\bibnamefont {Tumulka}}, \
  and\ \bibinfo {author} {\bibfnamefont {N.}~\bibnamefont {Zangh{\`{i}}}},\
  }\bibfield  {title} {\emph {\bibinfo {title} {{Canonical typicality}},\
  }}\href {\doibase 10.1103/PhysRevLett.96.050403} {\bibfield  {journal}
  {\bibinfo  {journal} {Phys. Rev. Lett.}\ }\textbf {\bibinfo {volume} {96}},\
  \bibinfo {pages} {050403} (\bibinfo {year} {2006})}\BibitemShut {NoStop}%
\bibitem [{\citenamefont {Popescu}\ \emph {et~al.}(2006)\citenamefont
  {Popescu}, \citenamefont {Short},\ and\ \citenamefont
  {Winter}}]{Popescu2006}%
  \BibitemOpen
  \bibfield  {author} {\bibinfo {author} {\bibfnamefont {S.}~\bibnamefont
  {Popescu}}, \bibinfo {author} {\bibfnamefont {A.~J.}\ \bibnamefont {Short}},
  \ and\ \bibinfo {author} {\bibfnamefont {A.}~\bibnamefont {Winter}},\
  }\bibfield  {title} {\emph {\bibinfo {title} {{Entanglement and the
  foundations of statistical mechanics}},\ }}\href {\doibase 10.1038/nphys444}
  {\bibfield  {journal} {\bibinfo  {journal} {Nat. Phys.}\ }\textbf {\bibinfo
  {volume} {2}},\ \bibinfo {pages} {754} (\bibinfo {year} {2006})}\BibitemShut
  {NoStop}%
\bibitem [{\citenamefont {Reimann}(2007)}]{Reimann2007}%
  \BibitemOpen
  \bibfield  {author} {\bibinfo {author} {\bibfnamefont {P.}~\bibnamefont
  {Reimann}},\ }\bibfield  {title} {\emph {\bibinfo {title} {{Typicality for
  generalized microcanonical ensembles}},\ }}\href
  {https://doi.org/10.1103/PhysRevLett.99.160404} {\bibfield  {journal}
  {\bibinfo  {journal} {Phys. Rev. Lett.}\ }\textbf {\bibinfo {volume} {99}},\
  \bibinfo {pages} {160404} (\bibinfo {year} {2007})}\BibitemShut {NoStop}%
\bibitem [{\citenamefont {Elsayed}\ and\ \citenamefont
  {Fine}(2013)}]{Elsayed2013}%
  \BibitemOpen
  \bibfield  {author} {\bibinfo {author} {\bibfnamefont {T.~A.}\ \bibnamefont
  {Elsayed}}\ and\ \bibinfo {author} {\bibfnamefont {B.~V.}\ \bibnamefont
  {Fine}},\ }\bibfield  {title} {\emph {\bibinfo {title} {{Regression relation
  for pure quantum states and its implications for efficient computing}},\
  }}\href {\doibase 10.1103/PhysRevLett.110.070404} {\bibfield  {journal}
  {\bibinfo  {journal} {Phys. Rev. Lett.}\ }\textbf {\bibinfo {volume} {110}},\
  \bibinfo {pages} {070404} (\bibinfo {year} {2013})}\BibitemShut {NoStop}%
\bibitem [{\citenamefont {Steinigeweg}\ \emph {et~al.}(2014)\citenamefont
  {Steinigeweg}, \citenamefont {Gemmer},\ and\ \citenamefont
  {Brenig}}]{Steinigeweg2014}%
  \BibitemOpen
  \bibfield  {author} {\bibinfo {author} {\bibfnamefont {R.}~\bibnamefont
  {Steinigeweg}}, \bibinfo {author} {\bibfnamefont {J.}~\bibnamefont {Gemmer}},
  \ and\ \bibinfo {author} {\bibfnamefont {W.}~\bibnamefont {Brenig}},\
  }\bibfield  {title} {\emph {\bibinfo {title} {{Spin-Current autocorrelations
  from single pure-state propagation}},\ }}\href
  {https://doi.org/10.1103/PhysRevLett.112.120601} {\bibfield  {journal}
  {\bibinfo  {journal} {Phys. Rev. Lett.}\ }\textbf {\bibinfo {volume} {112}},\
  \bibinfo {pages} {120601} (\bibinfo {year} {2014})}\BibitemShut {NoStop}%
\bibitem [{\citenamefont {Heitmann}\ \emph {et~al.}(2020)\citenamefont
  {Heitmann}, \citenamefont {Richter}, \citenamefont {Schubert},\ and\
  \citenamefont {Steinigeweg}}]{Heitmann2020}%
  \BibitemOpen
  \bibfield  {author} {\bibinfo {author} {\bibfnamefont {T.}~\bibnamefont
  {Heitmann}}, \bibinfo {author} {\bibfnamefont {J.}~\bibnamefont {Richter}},
  \bibinfo {author} {\bibfnamefont {D.}~\bibnamefont {Schubert}}, \ and\
  \bibinfo {author} {\bibfnamefont {R.}~\bibnamefont {Steinigeweg}},\
  }\bibfield  {title} {\emph {\bibinfo {title} {{Selected applications of
  typicality to real-time dynamics of quantum many-body systems}},\ }}\href
  {\doibase 10.1515/zna-2020-0010} {\bibfield  {journal} {\bibinfo  {journal}
  {Z. Naturforsch. A}\ }\textbf {\bibinfo {volume} {75}},\ \bibinfo {pages}
  {421} (\bibinfo {year} {2020})}\BibitemShut {NoStop}%
\bibitem [{\citenamefont {Jin}\ \emph {et~al.}(2021)\citenamefont {Jin},
  \citenamefont {Willsch}, \citenamefont {Willsch}, \citenamefont {Lagemann},
  \citenamefont {Michielsen},\ and\ \citenamefont {{De Raedt}}}]{Jin2021}%
  \BibitemOpen
  \bibfield  {author} {\bibinfo {author} {\bibfnamefont {F.}~\bibnamefont
  {Jin}}, \bibinfo {author} {\bibfnamefont {D.}~\bibnamefont {Willsch}},
  \bibinfo {author} {\bibfnamefont {M.}~\bibnamefont {Willsch}}, \bibinfo
  {author} {\bibfnamefont {H.}~\bibnamefont {Lagemann}}, \bibinfo {author}
  {\bibfnamefont {K.}~\bibnamefont {Michielsen}}, \ and\ \bibinfo {author}
  {\bibfnamefont {H.}~\bibnamefont {{De Raedt}}},\ }\bibfield  {title} {\emph
  {\bibinfo {title} {{Random state technology}},\ }}\href
  {https://doi.org/10.7566/JPSJ.90.012001} {\bibfield  {journal} {\bibinfo
  {journal} {J. Phys. Soc. Jpn.}\ }\textbf {\bibinfo {volume} {90}},\ \bibinfo
  {pages} {012001} (\bibinfo {year} {2021})}\BibitemShut {NoStop}%
\bibitem [{\citenamefont {Schollw\"ock}(2005)}]{Schollwoeck2005}%
  \BibitemOpen
  \bibfield  {author} {\bibinfo {author} {\bibfnamefont {U.}~\bibnamefont
  {Schollw\"ock}},\ }\bibfield  {title} {\emph {\bibinfo {title} {The
  density-matrix renormalization group},\ }}\href
  {https://link.aps.org/doi/10.1103/RevModPhys.77.259} {\bibfield  {journal}
  {\bibinfo  {journal} {Rev. Mod. Phys.}\ }\textbf {\bibinfo {volume} {77}},\
  \bibinfo {pages} {259} (\bibinfo {year} {2005})}\BibitemShut {NoStop}%
\bibitem [{\citenamefont {Schollwöck}(2011)}]{Schollwoeck2011}%
  \BibitemOpen
  \bibfield  {author} {\bibinfo {author} {\bibfnamefont {U.}~\bibnamefont
  {Schollwöck}},\ }\bibfield  {title} {\emph {\bibinfo {title} {The
  density-matrix renormalization group in the age of matrix product states},\
  }}\href {\doibase 10.1016/j.aop.2010.09.012} {\bibfield  {journal} {\bibinfo
  {journal} {Ann. Phys.}\ }\textbf {\bibinfo {volume} {326}},\ \bibinfo {pages}
  {96} (\bibinfo {year} {2011})}\BibitemShut {NoStop}%
\bibitem [{\citenamefont {Bertini}\ \emph {et~al.}(2021)\citenamefont
  {Bertini}, \citenamefont {Heidrich-Meisner}, \citenamefont {Karrasch},
  \citenamefont {Prosen}, \citenamefont {Steinigeweg},\ and\ \citenamefont
  {{\v{Z}}nidari{\v{c}}}}]{Bertini2021}%
  \BibitemOpen
  \bibfield  {author} {\bibinfo {author} {\bibfnamefont {B.}~\bibnamefont
  {Bertini}}, \bibinfo {author} {\bibfnamefont {F.}~\bibnamefont
  {Heidrich-Meisner}}, \bibinfo {author} {\bibfnamefont {C.}~\bibnamefont
  {Karrasch}}, \bibinfo {author} {\bibfnamefont {T.}~\bibnamefont {Prosen}},
  \bibinfo {author} {\bibfnamefont {R.}~\bibnamefont {Steinigeweg}}, \ and\
  \bibinfo {author} {\bibfnamefont {M.}~\bibnamefont {{\v{Z}}nidari{\v{c}}}},\
  }\bibfield  {title} {\emph {\bibinfo {title} {{Finite-temperature transport
  in one-dimensional quantum lattice models}},\ }}\href
  {https://doi.org/10.1103/RevModPhys.93.025003} {\bibfield  {journal}
  {\bibinfo  {journal} {Rev. Mod. Phys.}\ }\textbf {\bibinfo {volume} {93}},\
  \bibinfo {pages} {025003} (\bibinfo {year} {2021})}\BibitemShut {NoStop}%
\bibitem [{\citenamefont {Kubo}\ \emph {et~al.}(2012)\citenamefont {Kubo},
  \citenamefont {Toda},\ and\ \citenamefont {Hashisume}}]{Kubo2012}%
  \BibitemOpen
  \bibfield  {author} {\bibinfo {author} {\bibfnamefont {R.}~\bibnamefont
  {Kubo}}, \bibinfo {author} {\bibfnamefont {M.}~\bibnamefont {Toda}}, \ and\
  \bibinfo {author} {\bibfnamefont {N.}~\bibnamefont {Hashisume}},\ }\href
  {https://doi.org/10.1007/978-3-642-58244-8} {\emph {\bibinfo {title}
  {Statistical physics II: Nonequilibrium statistical mechanics}}}\ (\bibinfo
  {publisher} {Springer},\ \bibinfo {year} {2012})\BibitemShut {NoStop}%
\bibitem [{\citenamefont {Bastianello}\ \emph {et~al.}(2022)\citenamefont
  {Bastianello}, \citenamefont {Bertini}, \citenamefont {Doyon},\ and\
  \citenamefont {Vasseur}}]{Bastianello2022}%
  \BibitemOpen
  \bibfield  {author} {\bibinfo {author} {\bibfnamefont {A.}~\bibnamefont
  {Bastianello}}, \bibinfo {author} {\bibfnamefont {B.}~\bibnamefont
  {Bertini}}, \bibinfo {author} {\bibfnamefont {B.}~\bibnamefont {Doyon}}, \
  and\ \bibinfo {author} {\bibfnamefont {R.}~\bibnamefont {Vasseur}},\
  }\bibfield  {title} {\emph {\bibinfo {title} {Introduction to the special
  issue on emergent hydrodynamics in integrable many-body systems},\ }}\href
  {\doibase 10.1088/1742-5468/ac3e6a} {\bibfield  {journal} {\bibinfo
  {journal} {J. Stat. Mech. Theory Exp.}\ }\textbf {\bibinfo {volume} {2022}},\
  \bibinfo {pages} {014001} (\bibinfo {year} {2022})}\BibitemShut {NoStop}%
\bibitem [{\citenamefont {Doyon}\ \emph {et~al.}(2023)\citenamefont {Doyon},
  \citenamefont {Gopalakrishnan}, \citenamefont {Møller}, \citenamefont
  {Schmiedmayer},\ and\ \citenamefont {Vasseur}}]{Doyon2023}%
  \BibitemOpen
  \bibfield  {author} {\bibinfo {author} {\bibfnamefont {B.}~\bibnamefont
  {Doyon}}, \bibinfo {author} {\bibfnamefont {S.}~\bibnamefont
  {Gopalakrishnan}}, \bibinfo {author} {\bibfnamefont {F.}~\bibnamefont
  {Møller}}, \bibinfo {author} {\bibfnamefont {J.}~\bibnamefont
  {Schmiedmayer}}, \ and\ \bibinfo {author} {\bibfnamefont {R.}~\bibnamefont
  {Vasseur}},\ }\bibfield  {title} {\emph {\bibinfo {title} {Generalized
  hydrodynamics: a perspective},\ }}\href {https://arxiv.org/abs/2311.03438}
  {\bibfield  {journal} {\bibinfo  {journal} {arXiv:2311.03438}\ } (\bibinfo
  {year} {2023})}\BibitemShut {NoStop}%
\bibitem [{\citenamefont {Prosen}\ and\ \citenamefont
  {{\v{Z}}nidari{\v{c}}}(2009)}]{Prosen2009}%
  \BibitemOpen
  \bibfield  {author} {\bibinfo {author} {\bibfnamefont {T.}~\bibnamefont
  {Prosen}}\ and\ \bibinfo {author} {\bibfnamefont {M.}~\bibnamefont
  {{\v{Z}}nidari{\v{c}}}},\ }\bibfield  {title} {\emph {\bibinfo {title}
  {{Matrix product simulations of non-equilibrium steady states of quantum spin
  chains}},\ }}\href {\doibase 10.1088/1742-5468/2009/02/P02035} {\bibfield
  {journal} {\bibinfo  {journal} {J. Stat. Mech.}\ }\textbf {\bibinfo {volume}
  {2009}},\ \bibinfo {pages} {P02035} (\bibinfo {year} {2009})}\BibitemShut
  {NoStop}%
\bibitem [{\citenamefont {Michel}\ \emph {et~al.}(2003)\citenamefont {Michel},
  \citenamefont {Hartmann}, \citenamefont {Gemmer},\ and\ \citenamefont
  {Mahler}}]{Michel2003}%
  \BibitemOpen
  \bibfield  {author} {\bibinfo {author} {\bibfnamefont {M.}~\bibnamefont
  {Michel}}, \bibinfo {author} {\bibfnamefont {M.}~\bibnamefont {Hartmann}},
  \bibinfo {author} {\bibfnamefont {J.}~\bibnamefont {Gemmer}}, \ and\ \bibinfo
  {author} {\bibfnamefont {G.}~\bibnamefont {Mahler}},\ }\bibfield  {title}
  {\emph {\bibinfo {title} {{Fourier's Law confirmed for a class of small
  quantum systems}},\ }}\href {\doibase 10.1140/epjb/e2003-00228-x} {\bibfield
  {journal} {\bibinfo  {journal} {Eur. Phys. J. B}\ }\textbf {\bibinfo {volume}
  {34}},\ \bibinfo {pages} {325} (\bibinfo {year} {2003})}\BibitemShut
  {NoStop}%
\bibitem [{\citenamefont {Wichterich}\ \emph {et~al.}(2007)\citenamefont
  {Wichterich}, \citenamefont {Henrich}, \citenamefont {Breuer}, \citenamefont
  {Gemmer},\ and\ \citenamefont {Michel}}]{Wichterich2007}%
  \BibitemOpen
  \bibfield  {author} {\bibinfo {author} {\bibfnamefont {H.}~\bibnamefont
  {Wichterich}}, \bibinfo {author} {\bibfnamefont {M.~J.}\ \bibnamefont
  {Henrich}}, \bibinfo {author} {\bibfnamefont {H.-P.}\ \bibnamefont {Breuer}},
  \bibinfo {author} {\bibfnamefont {J.}~\bibnamefont {Gemmer}}, \ and\ \bibinfo
  {author} {\bibfnamefont {M.}~\bibnamefont {Michel}},\ }\bibfield  {title}
  {\emph {\bibinfo {title} {{Modeling heat transport through completely
  positive maps}},\ }}\href {\doibase 10.1103/PhysRevE.76.031115} {\bibfield
  {journal} {\bibinfo  {journal} {Phys. Rev. E}\ }\textbf {\bibinfo {volume}
  {76}},\ \bibinfo {pages} {031115} (\bibinfo {year} {2007})}\BibitemShut
  {NoStop}%
\bibitem [{\citenamefont {{\v{Z}}nidari{\v{c}}}(2011)}]{Znidaric2011}%
  \BibitemOpen
  \bibfield  {author} {\bibinfo {author} {\bibfnamefont {M.}~\bibnamefont
  {{\v{Z}}nidari{\v{c}}}},\ }\bibfield  {title} {\emph {\bibinfo {title} {{Spin
  transport in a one-dimensional anisotropic Heisenberg model}},\ }}\href
  {\doibase 10.1103/PhysRevLett.106.220601} {\bibfield  {journal} {\bibinfo
  {journal} {Phys. Rev. Lett.}\ }\textbf {\bibinfo {volume} {106}},\ \bibinfo
  {pages} {220601} (\bibinfo {year} {2011})}\BibitemShut {NoStop}%
\bibitem [{\citenamefont {Breuer}\ and\ \citenamefont
  {Petruccione}(2007)}]{Breuer2007}%
  \BibitemOpen
  \bibfield  {author} {\bibinfo {author} {\bibfnamefont {H.-P.}\ \bibnamefont
  {Breuer}}\ and\ \bibinfo {author} {\bibfnamefont {F.}~\bibnamefont
  {Petruccione}},\ }\href {\doibase 10.1093/acprof:oso/9780199213900.001.0001}
  {\emph {\bibinfo {title} {The theory of open quantum systems}}}\ (\bibinfo
  {publisher} {Oxford University Press},\ \bibinfo {year} {2007})\BibitemShut
  {NoStop}%
\bibitem [{\citenamefont {{De Raedt}}\ \emph {et~al.}(2017)\citenamefont {{De
  Raedt}}, \citenamefont {Jin}, \citenamefont {Katsnelson},\ and\ \citenamefont
  {Michielsen}}]{DeRaedt2017}%
  \BibitemOpen
  \bibfield  {author} {\bibinfo {author} {\bibfnamefont {H.}~\bibnamefont {{De
  Raedt}}}, \bibinfo {author} {\bibfnamefont {F.}~\bibnamefont {Jin}}, \bibinfo
  {author} {\bibfnamefont {M.~I.}\ \bibnamefont {Katsnelson}}, \ and\ \bibinfo
  {author} {\bibfnamefont {K.}~\bibnamefont {Michielsen}},\ }\bibfield  {title}
  {\emph {\bibinfo {title} {{Relaxation, thermalization, and Markovian dynamics
  of two spins coupled to a spin bath}},\ }}\href
  {https://doi.org/10.1103/PhysRevE.96.053306} {\bibfield  {journal} {\bibinfo
  {journal} {Phys. Rev. E}\ }\textbf {\bibinfo {volume} {96}},\ \bibinfo
  {pages} {053306} (\bibinfo {year} {2017})}\BibitemShut {NoStop}%
\bibitem [{\citenamefont {Verstraete}\ \emph {et~al.}(2004)\citenamefont
  {Verstraete}, \citenamefont {Garc{\'{i}}a-Ripoll},\ and\ \citenamefont
  {Cirac}}]{Verstraete2004}%
  \BibitemOpen
  \bibfield  {author} {\bibinfo {author} {\bibfnamefont {F.}~\bibnamefont
  {Verstraete}}, \bibinfo {author} {\bibfnamefont {J.~J.}\ \bibnamefont
  {Garc{\'{i}}a-Ripoll}}, \ and\ \bibinfo {author} {\bibfnamefont {J.~I.}\
  \bibnamefont {Cirac}},\ }\bibfield  {title} {\emph {\bibinfo {title} {{Matrix
  product density operators: Simulation of finite-temperature and dissipative
  systems}},\ }}\href {\doibase 10.1103/PhysRevLett.93.207204} {\bibfield
  {journal} {\bibinfo  {journal} {Phys. Rev. Lett.}\ }\textbf {\bibinfo
  {volume} {93}},\ \bibinfo {pages} {207204} (\bibinfo {year}
  {2004})}\BibitemShut {NoStop}%
\bibitem [{\citenamefont {Zwolak}\ and\ \citenamefont
  {Vidal}(2004)}]{Zwolak2004}%
  \BibitemOpen
  \bibfield  {author} {\bibinfo {author} {\bibfnamefont {M.}~\bibnamefont
  {Zwolak}}\ and\ \bibinfo {author} {\bibfnamefont {G.}~\bibnamefont {Vidal}},\
  }\bibfield  {title} {\emph {\bibinfo {title} {{Mixed-state dynamics in
  one-dimensional quantum lattice systems: A time-dependent superoperator
  renormalization algorithm}},\ }}\href
  {https://doi.org/10.1103/PhysRevLett.93.207205} {\bibfield  {journal}
  {\bibinfo  {journal} {Phys. Rev. Lett.}\ }\textbf {\bibinfo {volume} {93}},\
  \bibinfo {pages} {207205} (\bibinfo {year} {2004})}\BibitemShut {NoStop}%
\bibitem [{\citenamefont {Weimer}\ \emph {et~al.}(2021)\citenamefont {Weimer},
  \citenamefont {Kshetrimayum},\ and\ \citenamefont {Or{\'{u}}s}}]{Weimer2021}%
  \BibitemOpen
  \bibfield  {author} {\bibinfo {author} {\bibfnamefont {H.}~\bibnamefont
  {Weimer}}, \bibinfo {author} {\bibfnamefont {A.}~\bibnamefont
  {Kshetrimayum}}, \ and\ \bibinfo {author} {\bibfnamefont {R.}~\bibnamefont
  {Or{\'{u}}s}},\ }\bibfield  {title} {\emph {\bibinfo {title} {{Simulation
  methods for open quantum many-body systems}},\ }}\href
  {https://doi.org/10.1103/RevModPhys.93.015008} {\bibfield  {journal}
  {\bibinfo  {journal} {Rev. Mod. Phys.}\ }\textbf {\bibinfo {volume} {93}},\
  \bibinfo {pages} {015008} (\bibinfo {year} {2021})}\BibitemShut {NoStop}%
\bibitem [{\citenamefont {Prosen}(2011)}]{Prosen2011}%
  \BibitemOpen
  \bibfield  {author} {\bibinfo {author} {\bibfnamefont {T.}~\bibnamefont
  {Prosen}},\ }\bibfield  {title} {\emph {\bibinfo {title} {Open {$XXZ$} spin
  chain: Nonequilibrium steady state and a strict bound on ballistic
  transport},\ }}\href {\doibase 10.1103/PhysRevLett.106.217206} {\bibfield
  {journal} {\bibinfo  {journal} {Phys. Rev. Lett.}\ }\textbf {\bibinfo
  {volume} {106}},\ \bibinfo {pages} {217206} (\bibinfo {year}
  {2011})}\BibitemShut {NoStop}%
\bibitem [{\citenamefont {Prosen}\ and\ \citenamefont
  {Ilievski}(2013)}]{Prosen2013}%
  \BibitemOpen
  \bibfield  {author} {\bibinfo {author} {\bibfnamefont {T.}~\bibnamefont
  {Prosen}}\ and\ \bibinfo {author} {\bibfnamefont {E.}~\bibnamefont
  {Ilievski}},\ }\bibfield  {title} {\emph {\bibinfo {title} {Families of
  quasilocal conservation laws and quantum spin transport},\ }}\href
  {https://doi.org/10.1103/PhysRevLett.111.057203} {\bibfield  {journal}
  {\bibinfo  {journal} {Phys. Rev. Lett.}\ }\textbf {\bibinfo {volume} {111}},\
  \bibinfo {pages} {057203} (\bibinfo {year} {2013})}\BibitemShut {NoStop}%
\bibitem [{\citenamefont {Ljubotina}\ \emph {et~al.}(2019)\citenamefont
  {Ljubotina}, \citenamefont {{\v{Z}}nidari{\v{c}}},\ and\ \citenamefont
  {Prosen}}]{Ljubotina2019}%
  \BibitemOpen
  \bibfield  {author} {\bibinfo {author} {\bibfnamefont {M.}~\bibnamefont
  {Ljubotina}}, \bibinfo {author} {\bibfnamefont {M.}~\bibnamefont
  {{\v{Z}}nidari{\v{c}}}}, \ and\ \bibinfo {author} {\bibfnamefont
  {T.}~\bibnamefont {Prosen}},\ }\bibfield  {title} {\emph {\bibinfo {title}
  {{Kardar-Parisi-Zhang physics in the quantum Heisenberg magnet}},\ }}\href
  {\doibase 10.1103/PhysRevLett.122.210602} {\bibfield  {journal} {\bibinfo
  {journal} {Phys. Rev. Lett.}\ }\textbf {\bibinfo {volume} {122}},\ \bibinfo
  {pages} {210602} (\bibinfo {year} {2019})}\BibitemShut {NoStop}%
\bibitem [{\citenamefont {Gopalakrishnan}\ and\ \citenamefont
  {Vasseur}(2019)}]{Gopalakrishnan2019a}%
  \BibitemOpen
  \bibfield  {author} {\bibinfo {author} {\bibfnamefont {S.}~\bibnamefont
  {Gopalakrishnan}}\ and\ \bibinfo {author} {\bibfnamefont {R.}~\bibnamefont
  {Vasseur}},\ }\bibfield  {title} {\emph {\bibinfo {title} {{Kinetic theory of
  spin diffusion and superdiffusion in XXZ spin chains}},\ }}\href
  {https://doi.org/10.1103/PhysRevLett.122.127202} {\bibfield  {journal}
  {\bibinfo  {journal} {Phys. Rev. Lett.}\ }\textbf {\bibinfo {volume} {122}},\
  \bibinfo {pages} {127202} (\bibinfo {year} {2019})}\BibitemShut {NoStop}%
\bibitem [{\citenamefont {Bulchandani}\ \emph {et~al.}(2021)\citenamefont
  {Bulchandani}, \citenamefont {Gopalakrishnan},\ and\ \citenamefont
  {Ilievski}}]{Bulchandani2021}%
  \BibitemOpen
  \bibfield  {author} {\bibinfo {author} {\bibfnamefont {V.~B.}\ \bibnamefont
  {Bulchandani}}, \bibinfo {author} {\bibfnamefont {S.}~\bibnamefont
  {Gopalakrishnan}}, \ and\ \bibinfo {author} {\bibfnamefont {E.}~\bibnamefont
  {Ilievski}},\ }\bibfield  {title} {\emph {\bibinfo {title} {{Superdiffusion
  in spin chains}},\ }}\href {\doibase 10.1088/1742-5468/ac12c7} {\bibfield
  {journal} {\bibinfo  {journal} {J. Stat. Mech.}\ }\textbf {\bibinfo {volume}
  {2021}},\ \bibinfo {pages} {084001} (\bibinfo {year} {2021})}\BibitemShut
  {NoStop}%
\bibitem [{\citenamefont {Nandy}\ \emph {et~al.}(2023)\citenamefont {Nandy},
  \citenamefont {Lenar{\v{c}}i{\v{c}}}, \citenamefont {Ilievski}, \citenamefont
  {Mierzejewski}, \citenamefont {Herbrych},\ and\ \citenamefont
  {Prelov{\v{s}}ek}}]{Nandy2023}%
  \BibitemOpen
  \bibfield  {author} {\bibinfo {author} {\bibfnamefont {S.}~\bibnamefont
  {Nandy}}, \bibinfo {author} {\bibfnamefont {Z.}~\bibnamefont
  {Lenar{\v{c}}i{\v{c}}}}, \bibinfo {author} {\bibfnamefont {E.}~\bibnamefont
  {Ilievski}}, \bibinfo {author} {\bibfnamefont {M.}~\bibnamefont
  {Mierzejewski}}, \bibinfo {author} {\bibfnamefont {J.}~\bibnamefont
  {Herbrych}}, \ and\ \bibinfo {author} {\bibfnamefont {P.}~\bibnamefont
  {Prelov{\v{s}}ek}},\ }\bibfield  {title} {\emph {\bibinfo {title} {{Spin
  diffusion in a perturbed isotropic Heisenberg spin chain}},\ }}\href
  {\doibase 10.1103/PhysRevB.108.L081115} {\bibfield  {journal} {\bibinfo
  {journal} {Phys. Rev. B}\ }\textbf {\bibinfo {volume} {108}},\ \bibinfo
  {pages} {L081115} (\bibinfo {year} {2023})}\BibitemShut {NoStop}%
\bibitem [{\citenamefont {Serbyn}\ \emph {et~al.}(2021)\citenamefont {Serbyn},
  \citenamefont {Abanin},\ and\ \citenamefont {Papi{\'{c}}}}]{Serbyn2021}%
  \BibitemOpen
  \bibfield  {author} {\bibinfo {author} {\bibfnamefont {M.}~\bibnamefont
  {Serbyn}}, \bibinfo {author} {\bibfnamefont {D.~A.}\ \bibnamefont {Abanin}},
  \ and\ \bibinfo {author} {\bibfnamefont {Z.}~\bibnamefont {Papi{\'{c}}}},\
  }\bibfield  {title} {\emph {\bibinfo {title} {{Quantum many-body scars and
  weak breaking of ergodicity}},\ }}\href {\doibase 10.1038/s41567-021-01230-2}
  {\bibfield  {journal} {\bibinfo  {journal} {Nat. Phys.}\ }\textbf {\bibinfo
  {volume} {17}},\ \bibinfo {pages} {675} (\bibinfo {year} {2021})}\BibitemShut
  {NoStop}%
\bibitem [{\citenamefont {Singh}\ \emph {et~al.}(2021)\citenamefont {Singh},
  \citenamefont {Ware}, \citenamefont {Vasseur},\ and\ \citenamefont
  {Friedman}}]{Singh2021}%
  \BibitemOpen
  \bibfield  {author} {\bibinfo {author} {\bibfnamefont {H.}~\bibnamefont
  {Singh}}, \bibinfo {author} {\bibfnamefont {B.~A.}\ \bibnamefont {Ware}},
  \bibinfo {author} {\bibfnamefont {R.}~\bibnamefont {Vasseur}}, \ and\
  \bibinfo {author} {\bibfnamefont {A.~J.}\ \bibnamefont {Friedman}},\
  }\bibfield  {title} {\emph {\bibinfo {title} {{Subdiffusion and many-body
  quantum chaos with kinetic constraints}},\ }}\href
  {https://doi.org/10.1103/PhysRevLett.127.230602} {\bibfield  {journal}
  {\bibinfo  {journal} {Phys. Rev. Lett.}\ }\textbf {\bibinfo {volume} {127}},\
  \bibinfo {pages} {230602} (\bibinfo {year} {2021})}\BibitemShut {NoStop}%
\bibitem [{\citenamefont {Richter}\ and\ \citenamefont
  {Pal}(2022)}]{Richter2022a}%
  \BibitemOpen
  \bibfield  {author} {\bibinfo {author} {\bibfnamefont {J.}~\bibnamefont
  {Richter}}\ and\ \bibinfo {author} {\bibfnamefont {A.}~\bibnamefont {Pal}},\
  }\bibfield  {title} {\emph {\bibinfo {title} {{Anomalous hydrodynamics in a
  class of scarred frustration-free Hamiltonians}},\ }}\href {\doibase
  10.1103/PhysRevResearch.4.L012003} {\bibfield  {journal} {\bibinfo  {journal}
  {Phys. Rev. Research}\ }\textbf {\bibinfo {volume} {4}},\ \bibinfo {pages}
  {L012003} (\bibinfo {year} {2022})}\BibitemShut {NoStop}%
\bibitem [{\citenamefont {Richter}\ \emph {et~al.}(2023)\citenamefont
  {Richter}, \citenamefont {Lunt},\ and\ \citenamefont {Pal}}]{Richter2022}%
  \BibitemOpen
  \bibfield  {author} {\bibinfo {author} {\bibfnamefont {J.}~\bibnamefont
  {Richter}}, \bibinfo {author} {\bibfnamefont {O.}~\bibnamefont {Lunt}}, \
  and\ \bibinfo {author} {\bibfnamefont {A.}~\bibnamefont {Pal}},\ }\bibfield
  {title} {\emph {\bibinfo {title} {{Transport and entanglement growth in
  long-range random Clifford circuits}},\ }}\href
  {https://doi.org/10.1103/PhysRevResearch.5.L012031} {\bibfield  {journal}
  {\bibinfo  {journal} {Phys. Rev. Research}\ }\textbf {\bibinfo {volume}
  {5}},\ \bibinfo {pages} {L012031} (\bibinfo {year} {2023})}\BibitemShut
  {NoStop}%
\bibitem [{\citenamefont {Nandkishore}\ and\ \citenamefont
  {Huse}(2015)}]{Nandkishore2015}%
  \BibitemOpen
  \bibfield  {author} {\bibinfo {author} {\bibfnamefont {R.~M.}\ \bibnamefont
  {Nandkishore}}\ and\ \bibinfo {author} {\bibfnamefont {D.~A.}\ \bibnamefont
  {Huse}},\ }\bibfield  {title} {\emph {\bibinfo {title} {{Many-body
  localization and thermalization in quantum statistical mechanics}},\ }}\href
  {\doibase 10.1146/annurev-conmatphys-031214-014726} {\bibfield  {journal}
  {\bibinfo  {journal} {Annu. Rev. Condens. Matter Phys.}\ }\textbf {\bibinfo
  {volume} {6}},\ \bibinfo {pages} {15} (\bibinfo {year} {2015})}\BibitemShut
  {NoStop}%
\bibitem [{\citenamefont {Luitz}\ and\ \citenamefont {Lev}(2017)}]{Luitz2017a}%
  \BibitemOpen
  \bibfield  {author} {\bibinfo {author} {\bibfnamefont {D.~J.}\ \bibnamefont
  {Luitz}}\ and\ \bibinfo {author} {\bibfnamefont {Y.~B.}\ \bibnamefont
  {Lev}},\ }\bibfield  {title} {\emph {\bibinfo {title} {{The ergodic side of
  the many‐body localization transition}},\ }}\href
  {https://doi.org/10.1002/andp.201600350} {\bibfield  {journal} {\bibinfo
  {journal} {Ann. Phys.}\ }\textbf {\bibinfo {volume} {529}},\ \bibinfo {pages}
  {1600350} (\bibinfo {year} {2017})}\BibitemShut {NoStop}%
\bibitem [{\citenamefont {Steinigeweg}\ \emph
  {et~al.}(2009{\natexlab{a}})\citenamefont {Steinigeweg}, \citenamefont
  {Ogiewa},\ and\ \citenamefont {Gemmer}}]{Steinigeweg2009a}%
  \BibitemOpen
  \bibfield  {author} {\bibinfo {author} {\bibfnamefont {R.}~\bibnamefont
  {Steinigeweg}}, \bibinfo {author} {\bibfnamefont {M.}~\bibnamefont {Ogiewa}},
  \ and\ \bibinfo {author} {\bibfnamefont {J.}~\bibnamefont {Gemmer}},\
  }\bibfield  {title} {\emph {\bibinfo {title} {{Equivalence of transport
  coefficients in bath-induced and dynamical scenarios}},\ }}\href
  {https://doi.org/10.1209/0295-5075/87/10002} {\bibfield  {journal} {\bibinfo
  {journal} {EPL (Europhys. Lett.)}\ }\textbf {\bibinfo {volume} {87}},\
  \bibinfo {pages} {10002} (\bibinfo {year} {2009}{\natexlab{a}})}\BibitemShut
  {NoStop}%
\bibitem [{\citenamefont {Steinigeweg}\ and\ \citenamefont
  {Gemmer}(2009)}]{Steinigeweg2009c}%
  \BibitemOpen
  \bibfield  {author} {\bibinfo {author} {\bibfnamefont {R.}~\bibnamefont
  {Steinigeweg}}\ and\ \bibinfo {author} {\bibfnamefont {J.}~\bibnamefont
  {Gemmer}},\ }\bibfield  {title} {\emph {\bibinfo {title} {{Density dynamics
  in translationally invariant spin-1/2 chains at high temperatures: A
  current-autocorrelation approach}},\ }}\href
  {https://doi.org/10.1103/PhysRevB.80.184402} {\bibfield  {journal} {\bibinfo
  {journal} {Phys. Rev. B}\ }\textbf {\bibinfo {volume} {80}},\ \bibinfo
  {pages} {184402} (\bibinfo {year} {2009})}\BibitemShut {NoStop}%
\bibitem [{\citenamefont {{\v{Z}}nidari{\v{c}}}\ and\ \citenamefont
  {Ljubotina}(2018)}]{Znidaric2018}%
  \BibitemOpen
  \bibfield  {author} {\bibinfo {author} {\bibfnamefont {M.}~\bibnamefont
  {{\v{Z}}nidari{\v{c}}}}\ and\ \bibinfo {author} {\bibfnamefont
  {M.}~\bibnamefont {Ljubotina}},\ }\bibfield  {title} {\emph {\bibinfo {title}
  {{Interaction instability of localization in quasiperiodic systems}},\
  }}\href {\doibase 10.1073/pnas.1800589115} {\bibfield  {journal} {\bibinfo
  {journal} {Proc. Natl. Acad. Sci. USA}\ }\textbf {\bibinfo {volume} {115}},\
  \bibinfo {pages} {4595} (\bibinfo {year} {2018})}\BibitemShut {NoStop}%
\bibitem [{\citenamefont {{\v{Z}}nidari{\v{c}}}(2019)}]{Znidaric2019}%
  \BibitemOpen
  \bibfield  {author} {\bibinfo {author} {\bibfnamefont {M.}~\bibnamefont
  {{\v{Z}}nidari{\v{c}}}},\ }\bibfield  {title} {\emph {\bibinfo {title}
  {{Nonequilibrium steady-state Kubo formula: Equality of transport
  coefficients}},\ }}\href {\doibase 10.1103/PhysRevB.99.035143} {\bibfield
  {journal} {\bibinfo  {journal} {Phys. Rev. B}\ }\textbf {\bibinfo {volume}
  {99}},\ \bibinfo {pages} {035143} (\bibinfo {year} {2019})}\BibitemShut
  {NoStop}%
\bibitem [{\citenamefont {Kundu}\ \emph {et~al.}(2009)\citenamefont {Kundu},
  \citenamefont {Dhar},\ and\ \citenamefont {Narayan}}]{Kundu2009}%
  \BibitemOpen
  \bibfield  {author} {\bibinfo {author} {\bibfnamefont {A.}~\bibnamefont
  {Kundu}}, \bibinfo {author} {\bibfnamefont {A.}~\bibnamefont {Dhar}}, \ and\
  \bibinfo {author} {\bibfnamefont {O.}~\bibnamefont {Narayan}},\ }\bibfield
  {title} {\emph {\bibinfo {title} {{The Green–Kubo formula for heat
  conduction in open systems}},\ }}\href
  {https://doi.org/10.1088/1742-5468/2009/03/L03001} {\bibfield  {journal}
  {\bibinfo  {journal} {J. Stat. Mech.}\ }\textbf {\bibinfo {volume} {2009}},\
  \bibinfo {pages} {L03001} (\bibinfo {year} {2009})}\BibitemShut {NoStop}%
\bibitem [{\citenamefont {Purkayastha}\ \emph {et~al.}(2018)\citenamefont
  {Purkayastha}, \citenamefont {Sanyal}, \citenamefont {Dhar},\ and\
  \citenamefont {Kulkarni}}]{Purkayastha2018}%
  \BibitemOpen
  \bibfield  {author} {\bibinfo {author} {\bibfnamefont {A.}~\bibnamefont
  {Purkayastha}}, \bibinfo {author} {\bibfnamefont {S.}~\bibnamefont {Sanyal}},
  \bibinfo {author} {\bibfnamefont {A.}~\bibnamefont {Dhar}}, \ and\ \bibinfo
  {author} {\bibfnamefont {M.}~\bibnamefont {Kulkarni}},\ }\bibfield  {title}
  {\emph {\bibinfo {title} {{Anomalous transport in the
  Aubry-Andr{\'{e}}-Harper model in isolated and open systems}},\ }}\href
  {\doibase 10.1103/PhysRevB.97.174206} {\bibfield  {journal} {\bibinfo
  {journal} {Phys. Rev. B}\ }\textbf {\bibinfo {volume} {97}},\ \bibinfo
  {pages} {174206} (\bibinfo {year} {2018})}\BibitemShut {NoStop}%
\bibitem [{\citenamefont {Purkayastha}(2019)}]{Purkayastha2019}%
  \BibitemOpen
  \bibfield  {author} {\bibinfo {author} {\bibfnamefont {A.}~\bibnamefont
  {Purkayastha}},\ }\bibfield  {title} {\emph {\bibinfo {title} {{Classifying
  transport behavior via current fluctuations in open quantum systems}},\
  }}\href {\doibase 10.1088/1742-5468/ab02f4} {\bibfield  {journal} {\bibinfo
  {journal} {J. Stat. Mech.}\ }\textbf {\bibinfo {volume} {2019}},\ \bibinfo
  {pages} {043101} (\bibinfo {year} {2019})}\BibitemShut {NoStop}%
\bibitem [{\citenamefont {Heitmann}\ \emph
  {et~al.}(2023{\natexlab{a}})\citenamefont {Heitmann}, \citenamefont
  {Richter}, \citenamefont {Jin}, \citenamefont {Nandy}, \citenamefont
  {Lenar\ifmmode \check{c}\else \v{c}\fi{}i\ifmmode~\check{c}\else \v{c}\fi{}},
  \citenamefont {Herbrych}, \citenamefont {Michielsen}, \citenamefont
  {De~Raedt}, \citenamefont {Gemmer},\ and\ \citenamefont
  {Steinigeweg}}]{Heitmann2023}%
  \BibitemOpen
  \bibfield  {author} {\bibinfo {author} {\bibfnamefont {T.}~\bibnamefont
  {Heitmann}}, \bibinfo {author} {\bibfnamefont {J.}~\bibnamefont {Richter}},
  \bibinfo {author} {\bibfnamefont {F.}~\bibnamefont {Jin}}, \bibinfo {author}
  {\bibfnamefont {S.}~\bibnamefont {Nandy}}, \bibinfo {author} {\bibfnamefont
  {Z.}~\bibnamefont {Lenar\ifmmode \check{c}\else
  \v{c}\fi{}i\ifmmode~\check{c}\else \v{c}\fi{}}}, \bibinfo {author}
  {\bibfnamefont {J.}~\bibnamefont {Herbrych}}, \bibinfo {author}
  {\bibfnamefont {K.}~\bibnamefont {Michielsen}}, \bibinfo {author}
  {\bibfnamefont {H.}~\bibnamefont {De~Raedt}}, \bibinfo {author}
  {\bibfnamefont {J.}~\bibnamefont {Gemmer}}, \ and\ \bibinfo {author}
  {\bibfnamefont {R.}~\bibnamefont {Steinigeweg}},\ }\bibfield  {title} {\emph
  {\bibinfo {title} {Spin-$1/2$ {XXZ} chain coupled to two {L}indblad baths:
  Constructing nonequilibrium steady states from equilibrium correlation
  functions},\ }}\href {\doibase 10.1103/PhysRevB.108.L201119} {\bibfield
  {journal} {\bibinfo  {journal} {Phys. Rev. B}\ }\textbf {\bibinfo {volume}
  {108}},\ \bibinfo {pages} {L201119} (\bibinfo {year}
  {2023}{\natexlab{a}})}\BibitemShut {NoStop}%
\bibitem [{\citenamefont {Steinigeweg}\ \emph
  {et~al.}(2009{\natexlab{b}})\citenamefont {Steinigeweg}, \citenamefont
  {Wichterich},\ and\ \citenamefont {Gemmer}}]{Steinigeweg2009b}%
  \BibitemOpen
  \bibfield  {author} {\bibinfo {author} {\bibfnamefont {R.}~\bibnamefont
  {Steinigeweg}}, \bibinfo {author} {\bibfnamefont {H.}~\bibnamefont
  {Wichterich}}, \ and\ \bibinfo {author} {\bibfnamefont {J.}~\bibnamefont
  {Gemmer}},\ }\bibfield  {title} {\emph {\bibinfo {title} {{Density dynamics
  from current auto-correlations at finite time- and length-scales}},\ }}\href
  {\doibase 10.1209/0295-5075/88/10004} {\bibfield  {journal} {\bibinfo
  {journal} {EPL (Europhys. Lett.)}\ }\textbf {\bibinfo {volume} {88}},\
  \bibinfo {pages} {10004} (\bibinfo {year} {2009}{\natexlab{b}})}\BibitemShut
  {NoStop}%
\bibitem [{not({\natexlab{a}})}]{note1}%
  \BibitemOpen
  \href@noop {} {} ({\natexlab{a}}),\ \bibinfo {note} {the integrable case
  $\Delta' = 0$ is also diffusive for the anisotropy $\Delta = 1.5$
  \cite{Bertini2021}.}\BibitemShut {Stop}%
\bibitem [{\citenamefont {Steinigeweg}\ \emph {et~al.}(2017)\citenamefont
  {Steinigeweg}, \citenamefont {Jin}, \citenamefont {Schmidtke}, \citenamefont
  {{De Raedt}}, \citenamefont {Michielsen},\ and\ \citenamefont
  {Gemmer}}]{Steinigeweg2017a}%
  \BibitemOpen
  \bibfield  {author} {\bibinfo {author} {\bibfnamefont {R.}~\bibnamefont
  {Steinigeweg}}, \bibinfo {author} {\bibfnamefont {F.}~\bibnamefont {Jin}},
  \bibinfo {author} {\bibfnamefont {D.}~\bibnamefont {Schmidtke}}, \bibinfo
  {author} {\bibfnamefont {H.}~\bibnamefont {{De Raedt}}}, \bibinfo {author}
  {\bibfnamefont {K.}~\bibnamefont {Michielsen}}, \ and\ \bibinfo {author}
  {\bibfnamefont {J.}~\bibnamefont {Gemmer}},\ }\bibfield  {title} {\emph
  {\bibinfo {title} {{Real-time broadening of nonequilibrium density profiles
  and the role of the specific initial-state realization}},\ }}\href
  {https://doi.org/10.1103/PhysRevB.95.035155} {\bibfield  {journal} {\bibinfo
  {journal} {Phys. Rev. B}\ }\textbf {\bibinfo {volume} {95}},\ \bibinfo
  {pages} {035155} (\bibinfo {year} {2017})}\BibitemShut {NoStop}%
\bibitem [{\citenamefont {Dalibard}\ \emph {et~al.}(1992)\citenamefont
  {Dalibard}, \citenamefont {Castin},\ and\ \citenamefont
  {M{\o}lmer}}]{Dalibard1992}%
  \BibitemOpen
  \bibfield  {author} {\bibinfo {author} {\bibfnamefont {J.}~\bibnamefont
  {Dalibard}}, \bibinfo {author} {\bibfnamefont {Y.}~\bibnamefont {Castin}}, \
  and\ \bibinfo {author} {\bibfnamefont {K.}~\bibnamefont {M{\o}lmer}},\
  }\bibfield  {title} {\emph {\bibinfo {title} {{Wave-function approach to
  dissipative processes in quantum optics}},\ }}\href
  {https://doi.org/10.1103/PhysRevLett.68.580} {\bibfield  {journal} {\bibinfo
  {journal} {Phys. Rev. Lett.}\ }\textbf {\bibinfo {volume} {68}},\ \bibinfo
  {pages} {580} (\bibinfo {year} {1992})}\BibitemShut {NoStop}%
\bibitem [{\citenamefont {Michel}\ \emph {et~al.}(2008)\citenamefont {Michel},
  \citenamefont {Hess}, \citenamefont {Wichterich},\ and\ \citenamefont
  {Gemmer}}]{Michel2008}%
  \BibitemOpen
  \bibfield  {author} {\bibinfo {author} {\bibfnamefont {M.}~\bibnamefont
  {Michel}}, \bibinfo {author} {\bibfnamefont {O.}~\bibnamefont {Hess}},
  \bibinfo {author} {\bibfnamefont {H.}~\bibnamefont {Wichterich}}, \ and\
  \bibinfo {author} {\bibfnamefont {J.}~\bibnamefont {Gemmer}},\ }\bibfield
  {title} {\emph {\bibinfo {title} {{Transport in open spin chains: A Monte
  Carlo wave-function approach}},\ }}\href
  {https://doi.org/10.1103/PhysRevB.77.104303} {\bibfield  {journal} {\bibinfo
  {journal} {Phys. Rev. B}\ }\textbf {\bibinfo {volume} {77}},\ \bibinfo
  {pages} {104303} (\bibinfo {year} {2008})}\BibitemShut {NoStop}%
\bibitem [{\citenamefont {Heitmann}\ \emph
  {et~al.}(2023{\natexlab{b}})\citenamefont {Heitmann}, \citenamefont
  {Richter}, \citenamefont {Herbrych}, \citenamefont {Gemmer},\ and\
  \citenamefont {Steinigeweg}}]{Heitmann2022}%
  \BibitemOpen
  \bibfield  {author} {\bibinfo {author} {\bibfnamefont {T.}~\bibnamefont
  {Heitmann}}, \bibinfo {author} {\bibfnamefont {J.}~\bibnamefont {Richter}},
  \bibinfo {author} {\bibfnamefont {J.}~\bibnamefont {Herbrych}}, \bibinfo
  {author} {\bibfnamefont {J.}~\bibnamefont {Gemmer}}, \ and\ \bibinfo {author}
  {\bibfnamefont {R.}~\bibnamefont {Steinigeweg}},\ }\bibfield  {title} {\emph
  {\bibinfo {title} {{Real-time broadening of bath-induced density profiles
  from closed-system correlation functions}},\ }}\href
  {https://doi.org/10.1103/PhysRevE.108.024102} {\bibfield  {journal} {\bibinfo
   {journal} {Phys. Rev. E}\ }\textbf {\bibinfo {volume} {108}},\ \bibinfo
  {pages} {024102} (\bibinfo {year} {2023}{\natexlab{b}})}\BibitemShut
  {NoStop}%
\bibitem [{not({\natexlab{b}})}]{note2}%
  \BibitemOpen
  \href@noop {} {} ({\natexlab{b}}),\ \bibinfo {note} {while evaluating the
  prediction only requires averaging over jump times, the full stochastic
  unraveling also involves averaging over jump operators, which is the reason
  for slower convergence.}\BibitemShut {Stop}%
\bibitem [{\citenamefont {Wang}\ \emph {et~al.}(2023)\citenamefont {Wang},
  \citenamefont {Lamann}, \citenamefont {Steinigeweg},\ and\ \citenamefont
  {Gemmer}}]{Wang2023}%
  \BibitemOpen
  \bibfield  {author} {\bibinfo {author} {\bibfnamefont {J.}~\bibnamefont
  {Wang}}, \bibinfo {author} {\bibfnamefont {M.~H.}\ \bibnamefont {Lamann}},
  \bibinfo {author} {\bibfnamefont {R.}~\bibnamefont {Steinigeweg}}, \ and\
  \bibinfo {author} {\bibfnamefont {J.}~\bibnamefont {Gemmer}},\ }\bibfield
  {title} {\emph {\bibinfo {title} {Diffusion constants from the recursion
  method},\ }}\href {https://arxiv.org/abs/2312.02656} {\bibfield  {journal}
  {\bibinfo  {journal} {arXiv:2312.02656}\ } (\bibinfo {year}
  {2023})}\BibitemShut {NoStop}%
\bibitem [{\citenamefont {Richter}\ and\ \citenamefont
  {Steinigeweg}(2019)}]{Richter2019}%
  \BibitemOpen
  \bibfield  {author} {\bibinfo {author} {\bibfnamefont {J.}~\bibnamefont
  {Richter}}\ and\ \bibinfo {author} {\bibfnamefont {R.}~\bibnamefont
  {Steinigeweg}},\ }\bibfield  {title} {\emph {\bibinfo {title} {Combining
  dynamical quantum typicality and numerical linked cluster expansions},\
  }}\href {\doibase 10.1103/PhysRevB.99.094419} {\bibfield  {journal} {\bibinfo
   {journal} {Phys. Rev. B}\ }\textbf {\bibinfo {volume} {99}},\ \bibinfo
  {pages} {094419} (\bibinfo {year} {2019})}\BibitemShut {NoStop}%
\bibitem [{\citenamefont {Verstraete}\ \emph {et~al.}(2008)\citenamefont
  {Verstraete}, \citenamefont {Murg},\ and\ \citenamefont
  {Cirac}}]{Verstraete2008}%
  \BibitemOpen
  \bibfield  {author} {\bibinfo {author} {\bibfnamefont {F.}~\bibnamefont
  {Verstraete}}, \bibinfo {author} {\bibfnamefont {V.}~\bibnamefont {Murg}}, \
  and\ \bibinfo {author} {\bibfnamefont {J.~I.}\ \bibnamefont {Cirac}},\
  }\bibfield  {title} {\emph {\bibinfo {title} {Matrix product states,
  projected entangled pair states, and variational renormalization group
  methods for quantum spin systems},\ }}\href
  {https://doi.org/10.1080/14789940801912366} {\bibfield  {journal} {\bibinfo
  {journal} {Adv. Phys.}\ }\textbf {\bibinfo {volume} {57}},\ \bibinfo {pages}
  {143} (\bibinfo {year} {2008})}\BibitemShut {NoStop}%
\bibitem [{\citenamefont {Fendley}(2016)}]{Fendley2016}%
  \BibitemOpen
  \bibfield  {author} {\bibinfo {author} {\bibfnamefont {P.}~\bibnamefont
  {Fendley}},\ }\bibfield  {title} {\emph {\bibinfo {title} {{Strong zero modes
  and eigenstate phase transitions in the XYZ/interacting Majorana chain}},\
  }}\href {\doibase 10.1088/1751-8113/49/30/30LT01} {\bibfield  {journal}
  {\bibinfo  {journal} {J. Phys. A}\ }\textbf {\bibinfo {volume} {49}},\
  \bibinfo {pages} {30LT01} (\bibinfo {year} {2016})}\BibitemShut {NoStop}%
\bibitem [{\citenamefont {Kemp}\ \emph {et~al.}(2017)\citenamefont {Kemp},
  \citenamefont {Yao}, \citenamefont {Laumann},\ and\ \citenamefont
  {Fendley}}]{Kemp2017}%
  \BibitemOpen
  \bibfield  {author} {\bibinfo {author} {\bibfnamefont {J.}~\bibnamefont
  {Kemp}}, \bibinfo {author} {\bibfnamefont {N.~Y.}\ \bibnamefont {Yao}},
  \bibinfo {author} {\bibfnamefont {C.~R.}\ \bibnamefont {Laumann}}, \ and\
  \bibinfo {author} {\bibfnamefont {P.}~\bibnamefont {Fendley}},\ }\bibfield
  {title} {\emph {\bibinfo {title} {{Long coherence times for edge spins}},\
  }}\href {\doibase 10.1088/1742-5468/aa73f0} {\bibfield  {journal} {\bibinfo
  {journal} {J. Stat. Mech.}\ }\textbf {\bibinfo {volume} {2017}},\ \bibinfo
  {pages} {063105} (\bibinfo {year} {2017})}\BibitemShut {NoStop}%
\bibitem [{\citenamefont {Vasiloiu}\ \emph {et~al.}(2018)\citenamefont
  {Vasiloiu}, \citenamefont {Carollo},\ and\ \citenamefont
  {Garrahan}}]{Vasiloiu2018}%
  \BibitemOpen
  \bibfield  {author} {\bibinfo {author} {\bibfnamefont {L.~M.}\ \bibnamefont
  {Vasiloiu}}, \bibinfo {author} {\bibfnamefont {F.}~\bibnamefont {Carollo}}, \
  and\ \bibinfo {author} {\bibfnamefont {J.~P.}\ \bibnamefont {Garrahan}},\
  }\bibfield  {title} {\emph {\bibinfo {title} {Enhancing correlation times for
  edge spins through dissipation},\ }}\href {https://doi.org/10.1103/PhysRevB.98.094308}
  {\bibfield  {journal} {\bibinfo  {journal} {Phys.
  Rev. B}\ }\textbf {\bibinfo {volume} {98}},\ \bibinfo {pages} {094308}
  (\bibinfo {year} {2018})}\BibitemShut {NoStop}%
\end{thebibliography}

%

\clearpage

\newpage

\end{document}